\documentclass[11pt]{article}

\usepackage[T1]{fontenc}
\usepackage{lmodern,microtype}
\usepackage{inconsolata}
\usepackage{libertine}
\usepackage{amsmath,amssymb,amsthm}
\usepackage{fullpage,setspace}
\usepackage{enumitem,booktabs,needspace,placeins}
\usepackage[dvipsnames]{xcolor}
\usepackage{tikz}
\usetikzlibrary{arrows.meta,positioning,calc}
\usepackage{caption}
\usepackage{aliascnt}
\usepackage[breaklinks]{hyperref}
\hypersetup{
  colorlinks=true,
  linkcolor=blue,
  citecolor=blue,
  urlcolor=black
}
\usepackage[capitalise,nameinlink]{cleveref}
\usepackage[normalem]{ulem}

\newtheorem{theorem}{Theorem}[section]
\crefname{theorem}{Theorem}{Theorems}
\Crefname{theorem}{Theorem}{Theorems}

\newcommand{\DeclareSharedTheorem}[3]{%
  \newaliascnt{#1}{theorem}%
  \newtheorem{#1}[#1]{#2}%
  \aliascntresetthe{#1}%
  \crefname{#1}{#2}{#3}%
  \Crefname{#1}{#2}{#3}%
  \expandafter\def\csname #1autorefname\endcsname{#2}%
}

\DeclareSharedTheorem{lemma}{Lemma}{Lemmas}
\DeclareSharedTheorem{proposition}{Proposition}{Propositions}
\DeclareSharedTheorem{claim}{Claim}{Claims}
\DeclareSharedTheorem{fact}{Fact}{Facts}
\DeclareSharedTheorem{observation}{Observation}{Observations}
\DeclareSharedTheorem{corollary}{Corollary}{Corollaries}
\DeclareSharedTheorem{definition}{Definition}{Definitions}
\DeclareSharedTheorem{assumption}{Assumption}{Assumptions}
\DeclareSharedTheorem{remark}{Remark}{Remarks}

\crefname{invariant}{Invariant}{Invariants}
\Crefname{invariant}{Invariant}{Invariants}

\newcommand{\E}{\mathbb{E}}
\newcommand{\Prb}{\mathbb{P}}
\newcommand{\ALG}{\mathrm{ALG}}
\newcommand{\OPT}{\mathrm{OPT}}
\newcommand{\conn}{\operatorname{conn}}
\newcommand{\cost}{\operatorname{cost}}

\numberwithin{equation}{section}
\allowdisplaybreaks
\setlist{itemsep=3pt,topsep=5pt}
\AddToHook{env/theorem/before}{\Needspace{5\baselineskip}}
\AddToHook{env/lemma/before}{\Needspace{5\baselineskip}}
\AddToHook{env/proposition/before}{\Needspace{5\baselineskip}}
\AddToHook{env/corollary/before}{\Needspace{5\baselineskip}}

\usepackage{todonotes}

\title{Metric Facility Location under Batched Random Order}
\author{
Fabrizio Grandoni\thanks{IDSIA, USI-SUPSI, Switzerland, \textsf{fabrizio.grandoni@idsia.ch}. Work supported in part by the SNSF grant 200021-236706.}
\and
Stefano Leonardi\thanks{Sapienza University of Rome, Italy, \textsf{leonardi@diag.uniroma1.it}.}
\and
Matteo Russo\thanks{EPFL, Switzerland, \textsf{matteo.russo@epfl.ch}.}
}
\date{}

\begin{document}
\maketitle

\begin{abstract}
\noindent We study online metric facility location in the batched
random-order model. An adversary partitions $n$ clients into
$B$ nonempty batches, which arrive in uniformly random order.
Each batch is revealed in full, and its clients must be connected
to open facilities before the next batch arrives. Openings and
assignments are irrevocable, and the objective is to minimize
the total opening and connection cost.
For both uniform and nonuniform opening costs, we prove that
the tight competitive ratio is
\[
 \Theta\!\left(\min\left\{\log(B+1),
                         \sqrt{\log(n-B+2)}\right\}\right).
\]
Thus, the ratio is constant both when there is a single batch
and when every batch is a singleton, but intermediate numbers
of batches can require a $\Theta(\sqrt{\log n})$ ratio.
Our algorithms build on Meyerson's algorithm (2001) and do not
need to know $n$ nor $B$. A key algorithmic novelty is that we process the clients in a batch one by one in uniform random order (hence neglecting the global properties of the batch).
We present a novel analysis of this algorithm that hinges on the random order of the batches, as well as of the clients within each batch.
We also provide a matching competitive ratio lower bound
already for uniform costs, and even when the
entire metric is known in advance.
\end{abstract}

\section{Introduction}\label{sec:introduction}
Online metric facility location, introduced by
Meyerson~\cite{meyerson2001online}, is a cornerstone of online
algorithms and has attracted sustained
interest~\cite{Fotakis2008,KaplanNaoriRaz2023,
JiangEtAl2022,HuangJiang2026}.
Clients $v_1,\ldots,v_n$ from an underlying metric space $(X,d)$
arrive one by one, and each must be irrevocably assigned to an open facility upon arrival. Opening a facility at a permitted location $y$
costs $f(y)$, and assigning a client to it costs their distance.
The objective is to minimize the total opening and connection
cost. Every client location is a permitted facility location.
Opening costs are \emph{uniform} if $f(y)=f$ everywhere,
and \emph{nonuniform} otherwise. In the uniform case, our
algorithm opens facilities only at arriving clients.
In the nonuniform case, the permitted locations, their metric,
and their opening costs are known in advance.

The arrival order strongly affects the achievable guarantees.
A uniformly random permutation of the clients permits constant
competitive ratios for both opening-cost
models~\cite{meyerson2001online}, whereas adversarial order
already requires $\Theta(\log n/\log\log n)$ for uniform
costs~\cite{Fotakis2008}.
These assumptions describe two extremes: adversarial analysis
protects against every possible arrival sequence, while random
order assumes that all client permutations are equally likely.
Neither directly captures requests that arrive together because
they share a source or arise from the same event.

We study the \emph{batched random-order model}, introduced
for set cover by Gupta et al.~\cite{GuptaKehneLevin2021}. 
An oblivious adversary fixes the $n$ clients and partitions them
into $B$ nonempty batches, which arrive in uniformly random order. We remark that different batches might have different sizes.
Each batch is revealed in full, and all its clients must be
served before the next batch arrives.
 
This captures requests that arrive together because they share
a source or arise from the same event, while retaining randomness
in the order of these groups. For example, clients from one
region may arrive together, so an early batch need not represent
the full instance. Even after seeing that entire batch, the
algorithm must serve it without knowing the demand in other regions.

\begin{figure}[t]
\centering
\begingroup
\definecolor{batchblue}{HTML}{3266A8}
\definecolor{batchteal}{HTML}{258577}
\definecolor{batchpurple}{HTML}{8458A6}
\definecolor{batchgray}{HTML}{53616C}
\begin{tikzpicture}[
  x=1cm,y=1cm,
  font=\fontsize{9}{11}\selectfont,
  heading/.style={font=\fontsize{9}{11}\selectfont\bfseries,align=center},
  note/.style={font=\fontsize{8}{9.5}\selectfont,align=center,text=batchgray},
  flow/.style={-{Stealth[length=2mm]},line width=.65pt,draw=batchgray},
  client/.style={circle,inner sep=0pt,minimum size=4mm,
    fill=batchpurple,text=white,font=\fontsize{7}{8}\selectfont},
  batch/.style={rounded corners=3pt,line width=.8pt},
]
\node[heading] at (1.7,3.85) {Adversary fixes the batches};
\draw[rounded corners=5pt,draw=batchgray!50,densely dashed]
  (0,1.45) rectangle (3.4,3.5);
\draw[batch,draw=batchblue,fill=batchblue!6]
  (.20,2.55) rectangle (1.50,3.27);
\node[text=batchblue] at (.48,2.92) {$b_1$};
\fill[batchblue] (1.00,2.80) circle (2pt);
\fill[batchblue] (1.25,3.02) circle (2pt);
\draw[batch,draw=batchteal,fill=batchteal!6]
  (.20,1.68) rectangle (1.50,2.40);
\node[text=batchteal] at (.48,2.04) {$b_2$};
\foreach \x/\y in {.98/1.90,1.25/1.90,.98/2.17,1.25/2.17}
  \fill[batchteal] (\x,\y) circle (2pt);
\draw[batch,draw=batchpurple,fill=batchpurple!6]
  (1.72,1.97) rectangle (3.20,3.00);
\node[text=batchpurple] at (1.99,2.72) {$b_3$};
\node[client] at (2.44,2.70) {$a$};
\node[client] at (2.78,2.40) {$b$};
\node[client] at (2.27,2.27) {$c$};
\node[note] at (1.7,1.15) {Membership and sizes are fixed};
\draw[flow] (3.65,2.55) -- (5.45,2.55);
\node[note] at (4.55,3.03) {Uniform random\\batch order};
\node[heading] at (9.55,3.85) {Reveal each batch in full};
\draw[batch,draw=batchpurple,fill=batchpurple!6]
  (5.70,1.97) rectangle (7.70,3.20);
\node[text=batchpurple] at (5.99,2.92) {$b_3$};
\node[client] at (6.75,2.84) {$a$};
\node[client] at (7.15,2.47) {$b$};
\node[client] at (6.48,2.30) {$c$};
\draw[batch,draw=batchblue,fill=batchblue!6]
  (8.55,1.97) rectangle (10.55,3.20);
\node[text=batchblue] at (8.84,2.92) {$b_1$};
\fill[batchblue] (9.35,2.40) circle (2pt);
\fill[batchblue] (9.90,2.80) circle (2pt);
\draw[batch,draw=batchteal,fill=batchteal!6]
  (11.40,1.97) rectangle (13.40,3.20);
\node[text=batchteal] at (11.69,2.92) {$b_2$};
\foreach \x/\y in {12.20/2.35,12.72/2.35,12.20/2.82,12.72/2.82}
  \fill[batchteal] (\x,\y) circle (2pt);
\draw[flow] (7.82,2.55) -- (8.43,2.55);
\draw[flow] (10.67,2.55) -- (11.28,2.55);
\node[note] at (10.25,1.61) {Serve it before revealing the next};
\draw[flow,densely dashed] (6.70,1.95) -- (6.70,.97);
\draw[rounded corners=4pt,draw=batchgray!45,fill=batchgray!3]
  (5.70,-.02) rectangle (13.40,.97);
\node[client] (first) at (6.13,.48) {$c$};
\node[client] (second) at (6.70,.48) {$a$};
\node[client] (third) at (7.27,.48) {$b$};
\draw[flow] (first) -- (second);
\draw[flow] (second) -- (third);
\node[anchor=west,heading] at (7.80,.65)
  {Our algorithm: shuffle clients};
\node[anchor=west,note] at (7.80,.25)
  {Process one at a time within each batch};
\end{tikzpicture}
\endgroup
\caption{The adversary fixes the
clients and their batches, which are then uniformly permuted. Each batch is revealed in full, but our algorithms additionally shuffle its clients, as illustrated for batch $b_3$.}
\label{fig:batched-arrival-model}
\end{figure}
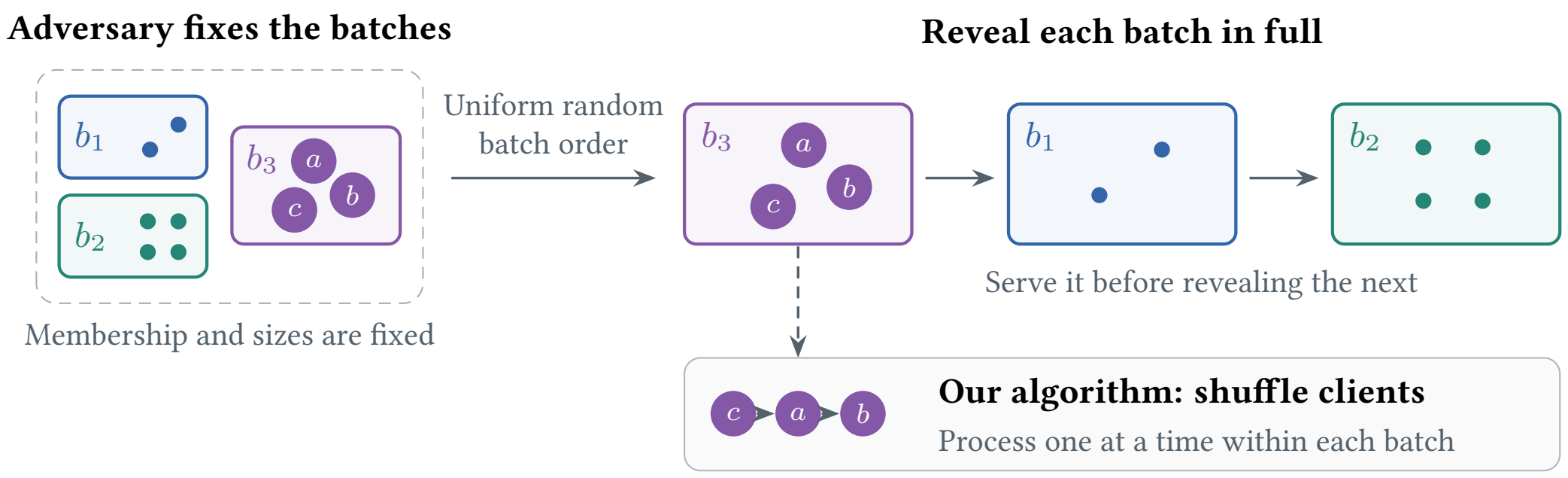

Two immediate upper bounds do not use the random batch order.
Solving each batch independently to within a constant factor
gives an $O(B)$-competitive algorithm, since the optimum for
any one batch costs at most the optimum for the full instance.
Alternatively, ignoring batch boundaries gives Meyerson's
$O(\log n)$ guarantee~\cite{meyerson2001online}, and, as mentioned, Fotakis~\cite{Fotakis2008}
sharpened the latter to the optimal
$\Theta(\log n/\log\log n)$ ratio for adversarial arrival order.
Moreover, revealing whole batches does not by itself permit
a sublinear dependence on $B$: the lower bound construction of ~\cite{Fotakis2008}
forces an $\Omega(B)$ lower bound when
the $B$ phases are revealed as batches of colocated clients.
Our question is how much the random order of the batches helps: we improve the dependence on $B$ from linear to logarithmic,
and the bound in terms of $n$ from $O(\log n/\log\log n)$
to $O(\sqrt{\log n})$.

\subsection{Our results and techniques}

Our main result determines the optimal competitive ratio in
the batched random-order model, with a tight dependence on both
the number of batches $B$ and the number of excess clients $n-B$.

\medskip
\noindent\textbf{Theorem (Informal version of
\Cref{thm:uniform-batched,thm:nonuniform-batched,lower:joint}).}
\emph{For $n$ clients in $B$ nonempty batches, the optimal
randomized competitive ratio, for both uniform and arbitrary
known positive opening costs, is}
\[
 \Theta\!\left(\min\left\{\log(B+1),
                         \sqrt{\log(n-B+2)}\right\}\right).
\]
In particular, the ratio is constant both when all clients arrive
together and when every batch is a singleton. For
$B=\Theta(n^\alpha)$ with fixed $0<\alpha<1$, it is instead
$\Theta(\sqrt{\log n})$.

Our algorithms shuffle uniformly at random
each batch and process its clients one at a time, although the
model allows joint decisions for the entire batch (\Cref{fig:batched-arrival-model}). Nonetheless, we obtain optimal asymptotic competitive ratios, which we found intriguing. We next describe the details of the algorithms for the uniform and non-uniform case.

\paragraph{Uniform costs.} We describe our results for the uniform opening cost case first.\medskip

\noindent $\rhd$ \textbf{The algorithm.}
We use Meyerson's algorithm~\cite{meyerson2001online}:
when client $v$ is processed, it opens a facility at $v$ with
probability $\min\{1,d(v,F_v)/f\}$, where $F_v$ is the set of
facilities already open, $f$ is their common opening cost, 
and $d(v,F_v)$ the distance of $v$ to its nearest facility in $F_v$. 
Otherwise, we connect $v$ to its nearest open facility.
Equivalently, for each client we draw an independent $U_v\sim\mathrm{Unif}(0,1)$ and open at $v$ if $fU_v\le d(v,F_v)$.
Following Meyerson, we describe the rule when clients can
themselves host facilities. If clients and permitted facility
sites are disjoint, we first map each client to its nearest
permitted site, which changes only the constant in the
connection-cost bound (\cref{obs:client-facility-reduction}), following Meyerson~\cite{meyerson2001online}.\medskip

We analyze one optimal cluster $C$ at a time, writing $c$ for
its facility and $\conn(C)=\sum_{v\in C}d(c,v)$.
Following the balanced-opening analysis of Huang and
Jiang~\cite{HuangJiang2026}, we identify a sufficient condition
for an opening to bring a facility closer to $c$; we call such
openings \emph{certified}. Their expected number controls the
algorithm's cost, up to an additive $O(\conn(C))$
(\cref{lem:cost-to-certified}).
Within a batch, the first certified client is biased towards
clients closer to $c$ and, therefore, its opening lets us charge
all later certifications in that batch to its clients'
connection cost (\cref{lem:within-batch}).
If $K_C$ counts batches containing a certification from $C$,
these two steps give
\[
 \E[\ALG(C)]\le O(\conn(C))+O(f)\E[K_C],
\]
where $\ALG(C)$ is the cost paid by our algorithm on cluster $C$.
Thus we need only to bound the number of batches that pay for
a first certification.\medskip

\noindent $\rhd$ \textbf{The logarithmic bound in $B$.} For this purpose, we use the algorithm's independent uniform
variables $U_v\in(0,1)$
 to assign the static keys
$Z_v=d(c,v)+fU_v$. After processing $v$, the distance to $c$,
capped at $f$, is strictly below $Z_v$.
Consequently, a batch containing a certification must have
a smaller minimum key than every preceding batch 
intersecting $C$.
Once the keys are fixed, the batches still arrive in uniformly
random order: the $r$th batch has the smallest key seen so far
with probability $1/r$. Summing gives
$\E[K_C]\le H_B = O(\log(B+1))$ (\cref{lem:records}).\medskip

\noindent $\rhd$ \textbf{The square-root logarithmic bound in $n-B$.}
We obtain a second bound on the same count by measuring how
much a certification decreases the distance to $c$.
Fix $\rho>1$. Starting from a distance capped at $f$, there
can be only $O(\log n/\log\rho)$ decreases by a factor
of $\rho$ before the capped distance falls below $f/n$.
After that, each remaining client is certified with probability
at most $1/n$, contributing at most one further certification
in expectation.
Every other certification requires a small lottery value,
namely $fU_v<(\rho-1)d(c,v)$.
The batches containing a client satisfying this condition are
fixed by the lottery variables, independently of the batch
order. Their expected number is at most
$(\rho-1)\conn(C)/f$, and we can count records among these
batches just as above. Together, the three cases give
(\cref{lem:records-II})
\begin{align*}
 \E[K_C]
 \le  O\!\left(1+\frac{\log n}{\log\rho}+\log\rho\right)
      +\frac{\conn(C)}{f},
\end{align*}
where the last term is absorbed into the connection cost.
Choosing $\log\rho=\max\{1,\sqrt{\log n}\}$ balances the
other two terms and gives an $O(\sqrt{\log n})$ bound.
To replace $n$ by $n-B$, we run separate copies on singleton
and nonsingleton batches, each using its own facility set.
The singleton clients arrive in
ordinary random order, where the algorithm is constant
competitive. The other copy receives at most
$\sum_{b:\,|b|\ge2}|b|\le 2(n-B)$ clients, and its batches retain uniformly random relative order.
We open the union of the two copies' facilities, so the actual
cost is at most the sum of their costs
(\cref{lem:joint-n-minus-B-transfer}).
The nonsingleton copy also retains the bound in $B$, giving
the minimum of the two guarantees for the combined algorithm.

\paragraph{Nonuniform costs.}
We round opening costs up to powers of two
$f_1<\cdots<f_m$ and use a marginal version of Meyerson's rule.
Let $\mathsf M_j$ contain the facilities costing at most $f_j$
after rounding. Before processing $v$, put
$\Delta_v^{(0)}=d(v,F_v)$ and
$\Delta_v^{(j)}=\min\{d(v,F_v),d(v,\mathsf M_j)\}$.
Independently at each level $j$, we select a nearest facility
in $\mathsf M_j$ for opening with probability
\[
 p_v^{(j)}=
 \min\left\{1,
 \frac{\Delta_v^{(j-1)}-\Delta_v^{(j)}}{f_j}\right\}.
\]
All 
lotteries for a client $v$
use the same facility set $F_v$, open before $v$ is considered.
We open the selected facilities and connect
$v$ to its nearest available facility.
Thus, each price level lottery pays for its marginal reduction in
distance to the client, rather than the entire distance
(\cref{eq:expanded-marginal-probability}).

The main new difficulty lies in the analysis within a batch.
At a single price level,
certification need not favor clients nearer to $c$, and
analyzing the price levels separately could charge to the algorithm the optimal connection costs within a cluster repeatedly for each level.
The key is to sum certification charges over all price levels before bounding connection costs in a way that consecutive marginal distances cancel. Within each batch, we distinguish between the half of clients closest to the optimal center and the half farthest from it. We then pair each client in the farther half with a distinct client in the nearer half. Either member of a pair arrives first with probability $1/2$, which lets us bound their total expected charge by a constant times the nearer client's charge (\cref{cl:far-clients}).

This allows us to bound separately the marginal cost incurred at each price level by using the same arguments used for the proof of the uniform case (\cref{nonunif:level-lemma}). We apply its record and progress bounds at each level and sum the resulting charges using the geometric prices, thus obtaining the same approximation bounds without introducing a factor for the number of levels (\cref{thm:nonuniform-batched}).

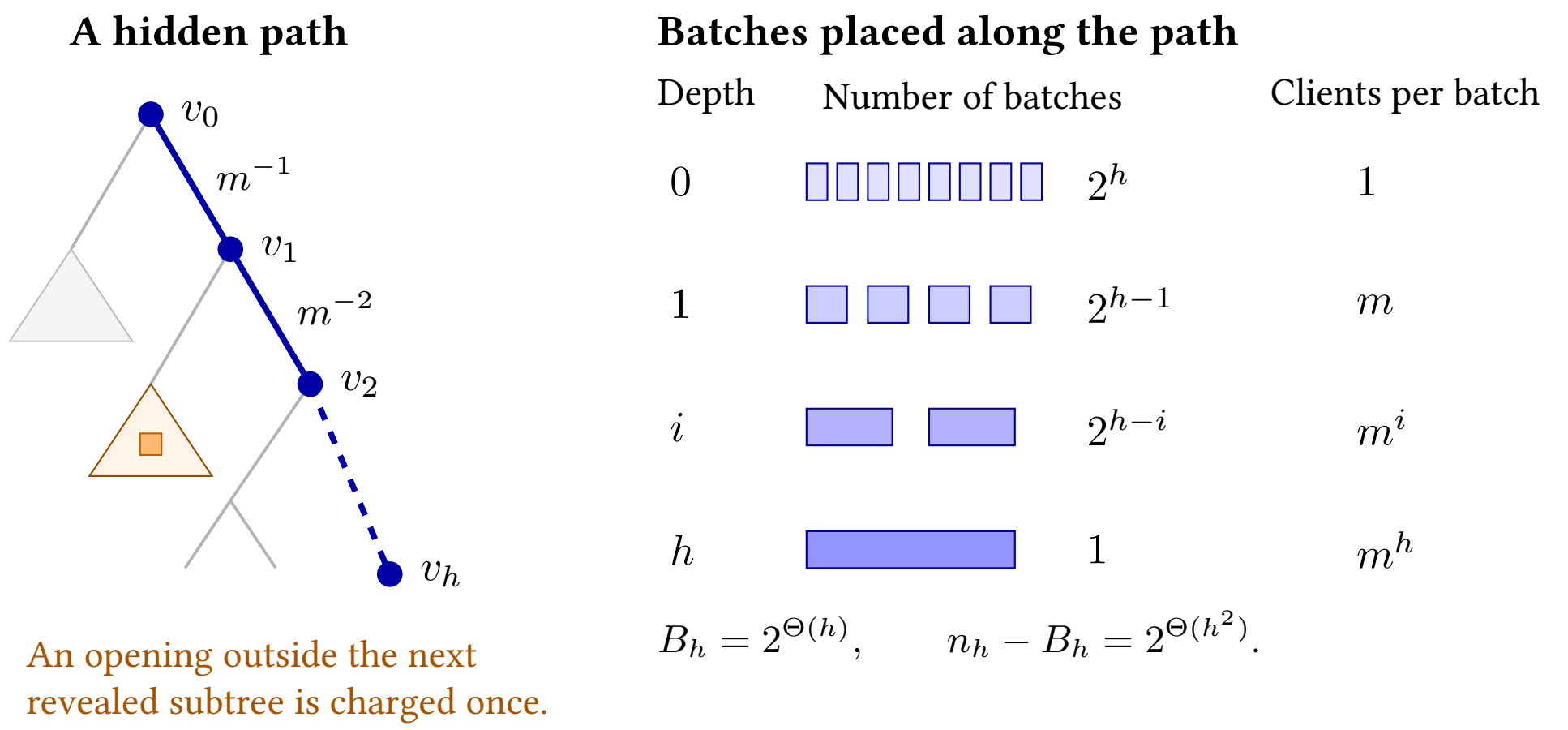
\begin{figure}[t]
\centering
\begin{tikzpicture}[x=1cm,y=1cm,font=\small,
  pathnode/.style={circle,fill=blue!65!black,inner sep=2.1pt},
  branch/.style={draw=black!30,line width=.7pt},
  hiddenpath/.style={draw=blue!65!black,line width=1.4pt}]
  \node[font=\small\bfseries,anchor=west] at (-.8,.65)
    {A hidden path};
  \coordinate (a) at (0,0);
  \coordinate (b) at (.65,-1.10);
  \coordinate (c) at (1.30,-2.20);
  \coordinate (d) at (1.95,-3.75);
  \draw[branch] (a)--(-.65,-1.10);
  \filldraw[fill=black!4,draw=black!25]
    (-1.15,-1.85)--(-.65,-1.10)--(-.15,-1.85)--cycle;
  \draw[branch] (b)--(0,-2.20);
  \filldraw[fill=orange!9,draw=orange!60!black]
    (-.50,-2.95)--(0,-2.20)--(.50,-2.95)--cycle;
  \node[draw=orange!75!black,fill=orange!55,
    rectangle,minimum size=5pt,inner sep=0pt] at (0,-2.69) {};
  \draw[branch] (c)--(.65,-3.15);
  \draw[branch] (.65,-3.15)--(.28,-3.70);
  \draw[branch] (.65,-3.15)--(1.02,-3.70);
  \draw[hiddenpath] (a)--(b);
  \draw[hiddenpath] (b)--(c);
  \draw[hiddenpath,dashed] (c)--(d);
  \node[pathnode,label=right:{$v_0$}] at (a) {};
  \node[pathnode,label=right:{$v_1$}] at (b) {};
  \node[pathnode,label=right:{$v_2$}] at (c) {};
  \node[pathnode,label=right:{$v_h$}] at (d) {};
  \node[anchor=west,font=\footnotesize,fill=white,inner sep=1pt]
    at (.49,-.48) {$m^{-1}$};
  \node[anchor=west,font=\footnotesize,fill=white,inner sep=1pt]
    at (1.15,-1.58) {$m^{-2}$};
  \node[anchor=north west,align=left,font=\footnotesize,
    text=orange!65!black] at (-1.15,-4.15)
    {An opening outside the next\\revealed subtree is charged once.};

  \node[font=\small\bfseries,anchor=west] at (4.0,.65)
    {Batches placed along the path};
  \node[anchor=west,font=\footnotesize] at (4.0,.15) {Depth};
  \node[anchor=west,font=\footnotesize] at (5.35,.15) {Number of batches};
  \node[anchor=west,font=\footnotesize] at (9.0,.15) {Clients per batch};
  \foreach \y/\dep/\cnt/\siz in
    {-.55/0/{2^h}/1,-1.55/1/{2^{h-1}}/m,
     -2.55/i/{2^{h-i}}/{m^i},-3.55/h/1/{m^h}} {
    \node[anchor=west] at (4.1,\y) {$\dep$};
    \node[anchor=west] at (7.5,\y) {$\cnt$};
    \node[anchor=west] at (9.7,\y) {$\siz$};
  }
  \foreach \x in {5.35,5.60,5.85,6.10,6.35,6.60,6.85,7.10}
    \draw[blue!65!black,fill=blue!12] (\x,-.70) rectangle +(0.17,.30);
  \foreach \x in {5.35,5.85,6.35,6.85}
    \draw[blue!65!black,fill=blue!20] (\x,-1.70) rectangle +(0.33,.30);
  \foreach \x in {5.35,6.35}
    \draw[blue!65!black,fill=blue!30] (\x,-2.70) rectangle +(0.70,.30);
  \draw[blue!65!black,fill=blue!42] (5.35,-3.70) rectangle +(1.70,.30);
  \node[anchor=west,font=\footnotesize] at (4.0,-4.25)
    {$B_h=2^{\Theta(h)},\qquad n_h-B_h=2^{\Theta(h^2)}$.};
\end{tikzpicture}
\caption{The lower-bound construction, with $m=2^{h+3}$.
The blue path is chosen at random and hidden from the algorithm;
the gray branches show other possible continuations.
We place $2^{h-i}$ batches of $m^i$ clients at depth $i$.
Having more batches at shallower depths ensures that each depth
is revealed before all deeper ones with probability at least $1/2$.
The edge entering a depth-$i$ subtree has length $m^{-i}$,
so serving one of its batches entirely from outside that subtree
costs at least $m^i m^{-i}=1$.
In contrast, the offline solution can serve all clients from
a single facility at the final leaf for a total cost of $O(1)$.}
\label{fig:intro-batched-lower-bound}
\end{figure}

\paragraph{Matching lower bounds.}
We adapt Fotakis's hidden-path tree
construction~\cite{Fotakis2008} (\cref{fig:intro-batched-lower-bound}).
The metric is a known binary tree of depth $h$, with unit
opening costs, but all clients lie on a uniformly chosen
root-to-leaf path that is hidden from the algorithm.
A newly revealed vertex deeper than all previous ones
narrows the possible continuation of the path.
The batch sizes ensure that serving a batch from outside
its subtree costs at least one. Alternatively, facilities
opened inside the current subtree may be left outside the
next revealed subtree, and each such opening can be charged
only once. Together, these charges give an expected cost
proportional to the number of new deepest revelations
(\cref{lower:record-charge}).

The obstacle is that with one batch per depth level, random order
produces only $H_{h+1}=O(\log h)$ new deepest revelations,
rather than $\Theta(h)$. We therefore place $2^{h-i}$ batches
at depth $i$. This depth is revealed before every deeper one
with probability
\[
 \frac{2^{h-i}}{\sum_{\ell=i}^h2^{h-\ell}}
 =\frac{2^{h-i}}{2^{h-i+1}-1}\ge\frac12,
\]
so the expected online cost is now $\Omega(h)$.
The number of batches is $B_h=2^{h+1}-1$, explaining the
logarithmic dependence on $B$.
To keep the offline cost small despite these repetitions,
we shrink edge lengths by a factor $m=2^{h+3}$ at each depth.
A batch at depth $i$ contains $m^i$ clients, and its parent
edge has length $m^{-i}$.
Opening just at the hidden leaf gives
\[
 \OPT\le1+\sum_{i=0}^{h-1}
       2^{h-i}m^i\sum_{\ell=i+1}^h m^{-\ell}
 \le1+\frac{2^{h+1}-2}{m-1}<\frac54.
\]
However, these large batches use
$n_h=\sum_{i=0}^h2^{h-i}m^i=\exp(\Theta(h^2))$
clients, and hence $\log(n_h-B_h)=\Theta(h^2)$
(\cref{lower:construction}).
Thus depth $h$ costs exponentially many batches in $h$ and
exponentially many excess clients in $h^2$.
Choosing the largest depth that fits both budgets yields
the two terms in our lower bound.
Finally, adding singleton batches and extra clients at the
root gives every feasible pair $(n,B)$: one additional root
facility serves all added clients without revealing more of
the hidden path (\cref{lower:joint}).

\subsection{Related work}

\paragraph{Metric facility location.} Offline metric facility location has been a testing ground for approximation
techniques. Shmoys et al.~\cite{ShmoysTardosAardal1997} gave
the first constant-factor approximation by rounding a linear-programming
relaxation. Mahdian et al.~\cite{MahdianYeZhang2006} obtained a
$1.52$-approximation using cost scaling, greedy improvement, and dual
fitting; Li~\cite{Li2013} later obtained a $1.488$-approximation by
randomizing an LP-rounding parameter and combining the resulting algorithm
with an earlier greedy algorithm.
A related line of work seeks stronger guarantees that preserve the
coefficient of the opening cost while approximating the connection cost.
Cohen-Addad et al.~\cite{CohenAddadEtAl2023} combined dual fitting
and local search to obtain a Lagrangian-multiplier-preserving approximation
strictly below $2$, with consequences for $k$-median. This guarantee is
different from the usual approximation ratio for the total cost.

\paragraph{Metric facility location with general cost functions.} Offline variants also connect facility location to network design and
more general cost functions. Leonardi and
Sch\"afer~\cite{LeonardiSchaefer2004} studied cost sharing for connected
facility location, where the open facilities must also be joined by a
Steiner tree. They developed methods that recover a constant fraction of
the solution cost while ensuring that a client's charge cannot increase
when more clients participate, including a primal-dual method with
nonzero opening costs.
For the related Steiner-tree problem, Byrka et al.~\cite{ByrkaEtAl2013} gave a $(\ln 4+\varepsilon)$-approximation
using iterative randomized rounding; Steiner-tree algorithms serve as
building blocks in connected facility location.
Abbasi et al.~\cite{AbbasiEtAl2024} considered submodular facility
location, where a facility's cost depends on its assigned set of clients
through a monotone submodular function, and obtained an
$O(\log\log n)$-approximation.
These works address computation when all clients are known. Our bounds
instead measure the cost of making irrevocable decisions before future
batches are revealed.

\paragraph{Online metric facility location.} Meyerson~\cite{meyerson2001online} introduced online facility location and
proved constant competitiveness in random order for uniform and nonuniform
opening costs. Fotakis~\cite{Fotakis2008} established the optimal adversarial
ratio $\Theta(\log n/\log\log n)$.
For uniform costs, Kaplan et al.~\cite{KaplanNaoriRaz2023} proved
that Meyerson's algorithm is exactly $4$-competitive in random order and
gave a $3$-competitive modification.
Recent work further improves the constants using arrival times: Huang and
Jiang~\cite{HuangJiang2026} obtain a ratio below $2.42$ for uniform costs,
while Peng and Tang~\cite{PengTang2026} improve the nonuniform-cost bounds.
Huang and Jiang also use the center-excess certification
$(d(c,F)-d(c,v))_+$ that underlies our uniform analysis.
Our contribution is its batchwise analysis through static keys and progress
bounds, and the extensions to nonuniform costs.

\paragraph{Learning-augmented online metric facility location.} Predictions offer another way to improve online facility location.
Jiang et al.~\cite{JiangEtAl2022} augment the algorithm with predictions of
the optimal facility serving each client and analyze the dependence on the
maximum prediction error.
For uniform costs, Fotakis et al.~\cite{FotakisEtAl2025} obtain matching
bounds in terms of the sum of prediction errors.
Lattanzi et al.~\cite{LattanziPanigrahiSvensson2026}
develop online rounding algorithms and applications to learning-augmented
facility location. These works exploit advice about future requests or an
optimal solution; our algorithms instead exploit the arrival distribution.

\paragraph{Algorithms in the random-order model.} The random-order model has also been studied extensively for secretary,
packing, and covering problems; Gupta and Singla~\cite{GuptaSingla2021}
survey the principal techniques and their relation to other stochastic
models. For facility location, Lang~\cite{Lang2018} studies an interpolation
in which an adversary can buffer and reorder a bounded number of requests
from an initially random stream. Kaplan et al.~\cite{KaplanNaoriRaz2023}
also study partial adversarial orders. These models alter a random stream,
whereas here a fixed partition or hierarchy determines which groups are
permuted. For set cover with $m$ sets and $n$ elements, Gupta et al.~\cite{GuptaKehneLevin2021} give an $O(\log(mn))$-competitive
random-order algorithm, \emph{LearnOrCover}.
The same paper introduces batched random-order set cover: entire batches
are revealed in a uniformly random order. For $b$ equal-size batches of
$s$ elements, they prove a lower bound of $\Omega(\log b\log s)$ for
polynomial-time randomized algorithms, unless $\mathrm{NP}\subseteq
\mathrm{BPP}$.
We depart slightly from their batch-revelation convention, as our algorithms do not need batches to be revealed in full, and the batches need not be of the same size. However, our lower bounds allow every use of the fully revealed batch which makes the problem only easier.
Gupta et al.~\cite{GuptaKehneLevin2024} extend the covering
framework to nonmetric facility location and relate random-order algorithms
to single-sample prophet algorithms.
Gupta et al.~\cite{GuptaMolinaroRusso2026} give a direct
connection to stochastic mirror descent and online convex optimization,
recovering the set-cover guarantee and extensions to other covering
problems. In the metric setting considered here, the triangle inequality
allows us to compare clients served by the same optimal facility and to
measure how an opening helps future clients in that cluster.

\section{Preliminaries}\label{sec:prelim}

In the \emph{batched random-order} metric facility location problem, an oblivious adversary partitions a multiset $V = \{v_1, \ldots, v_n\}$ of $n$ clients into
$B$ nonempty batches $\{b_1, \ldots, b_B\}$ not necessarily of equal size. Batches arrive in uniformly random order, and each batch is revealed in full and must be served before
the next arrives. Our algorithms independently shuffle each
batch and process its clients one at a time.

\paragraph{Notation.} We write $d(v,F)=\min_{y\in F}d(v,y)$, with
$d(v,\varnothing)=\infty$, $[x]_+=\max\{x,0\}$, and
$H_s=\sum_{i=1}^s1/i$. Fixed an optimal solution with clusters $\mathcal C$, for an optimal cluster $C$ served by facility $c$, write $C_b=C\cap b$
and $\conn(S)=\sum_{v\in S}d(c,v)$ for $S\subseteq C$.
Thus, the offline optimal cost can be decomposed into the optimal connection cost and optimal opening cost: $\OPT_{\rm conn}=\sum_C\conn(C)$,
$\OPT_{\rm open}=\sum_C f(c)$, and
$\OPT=\OPT_{\rm conn}+\OPT_{\rm open}$.
Similarly, if we let $\cost(v)$ be the cost incurred while processing $v$, and
$\ALG(C)=\sum_{v\in C}\cost(v)$, then the total algorithm's cost is $\ALG=\sum_C\ALG(C)$. Finally, as it is customary, an algorithm is said to be $\alpha$-competitive if $\E[\ALG]\le \alpha\OPT$,
where $\alpha \ge 1$.

\paragraph{Clients and facility sites.}
As mentioned in \Cref{sec:introduction}, for uniform costs, we follow Meyerson~\cite{meyerson2001online}
in presenting the algorithm with openings at client locations.

\begin{observation}\label{obs:client-facility-reduction}
Any bound $a\OPT_{\rm conn}+b\OPT_{\rm open}$ valid against
every fixed comparison solution when clients are permitted
facility sites extends to
\[
 \E[\ALG]\le(2a+1)\OPT_{\rm conn}+b\OPT_{\rm open}
\]
for arbitrary client locations and known permitted sites.
This holds for uniform and nonuniform costs, preserving
$n$, $B$, and the arrival distribution.
\end{observation}
\begin{proof}
Map each client $v$ to its nearest permitted site $y(v)$,
retaining its occurrence and batch, and run the algorithm
with unchanged facility costs. For every comparison facility $c$,
\[
 d(v,y(v))\le d(v,c),
 \qquad d(y(v),c)\le2d(v,c).
\]
The original optimum therefore serves the projected clients
with unchanged opening cost and at most twice its connection
cost. Connecting each original client to the facility serving
its projection adds at most
$\sum_v d(v,y(v))\le\OPT_{\rm conn}$, proving the bound.
\end{proof}

\paragraph{Transferring bounds from $n$ to $n-B$.}
Our bounds depend on
both $B$ and $n-B$. Our analysis first gives bounds in terms
of $B$ and the total number of clients $n$. The following
observation sharpens the dependence on $n$ to one on $n-B$:
singleton batches can be handled at constant competitive ratio,
while nonsingleton batches contain at most $2(n-B)$ clients.

\begin{observation}[Singleton--nonsingleton transfer]
\label{lem:joint-n-minus-B-transfer}\label{unif:transfer}
Suppose a batched random-order algorithm satisfies
\[
 \E[\ALG]\le a\OPT_{\rm conn}
       +\mathrm{CR}_{\mathrm{BR}}(m)\OPT_{\rm open}
\]
on instances with at most $m$ clients, where $a$ is a constant
and $\mathrm{CR}_{\mathrm{BR}}$ is nondecreasing.
Assume the same bound holds against every fixed feasible
comparison solution.
If an ordinary random-order algorithm in the same opening-cost
model has competitive ratio $\mathrm{CR}_{\mathrm{RO}}$,
then the cost $\ALG_{\mathrm{comb}}$ of the combined algorithm satisfies
\[
 \E[\ALG_{\mathrm{comb}}]
 \le \max\{a,\mathrm{CR}_{\mathrm{RO}}\}\OPT_{\rm conn}
 +\bigl(\mathrm{CR}_{\mathrm{RO}}
       +\mathrm{CR}_{\mathrm{BR}}(2(n-B))\bigr)\OPT_{\rm open}.
\]
\end{observation}

\begin{proof}
Run the ordinary random-order algorithm on singleton batches
and the batched algorithm on nonsingleton batches.
Each copy uses its own facilities for its decisions; the
combined algorithm opens their union and connects each client
to the closest facility in that union.
Both subsequences retain uniformly random relative order,
and the nonsingleton copy receives at most $\sum_{b:|b|\ge2}|b| \le2\sum_{b:|b|\ge2}(|b|-1)=2(n-B)$ clients. We next compare both copies with the same optimum restricted to their
clients. The restricted connection costs sum to
$\OPT_{\rm conn}$, and each restricted opening cost is at most
$\OPT_{\rm open}$.
et $\ALG_{\mathrm{sing}}$ and $\ALG_{\mathrm{nonsing}}$
be the costs paid by the two copies on the singleton and
nonsingleton batches, respectively.
Since the combined cost is at most the sum
of the copies' costs,
\begin{align*}
 \E[\ALG_{\mathrm{comb}}]
 &\le\E[\ALG_{\mathrm{sing}}]+\E[\ALG_{\mathrm{nonsing}}]\\
 &\le \mathrm{CR}_{\mathrm{RO}}
       \sum_{C, \, b:\,|b|=1}\conn(C_b)
      +a \sum_{C, \, b:\,|b|\ge2}\conn(C_b) +\bigl(\mathrm{CR}_{\mathrm{RO}}
       +\mathrm{CR}_{\mathrm{BR}}(2(n-B))\bigr)\OPT_{\rm open}\\
 &\le\max\{a,\mathrm{CR}_{\mathrm{RO}}\}\OPT_{\rm conn}
      +\bigl(\mathrm{CR}_{\mathrm{RO}}
       +\mathrm{CR}_{\mathrm{BR}}(2(n-B))\bigr)\OPT_{\rm open}.
\end{align*}
The last inequality uses
$\sum_{C, \,b:\,|b|=1}\conn(C_b)
+\sum_{C, \, b:\,|b|\ge2}\conn(C_b)=\OPT_{\rm conn}$:
each client's connection cost appears in exactly one of the
two sums, so we take the maximum of their coefficients.
\end{proof}

\section{Uniform opening costs}
\label{sec:uniform-costs}\label{sec:uniform}

Let $f>0$ be the common opening cost. We first reduce the
algorithm's cost to certified openings, then to one charge per
batch. A record argument bounds the number of these charges;
a second bound combines records with decreases in the distance
to an optimal center.

\begin{theorem}[Uniform opening costs]
\label{thm:uniform-batched}\label{unif:main}
For $n\ge1$ clients in $1\le B\le n$ batches, there exists an algorithm satisfying
\[
 \E[\ALG]\le O(1)\OPT_{\mathrm{conn}}
 +O\!\left(\min\left\{\log(B+1),\sqrt{\log(n-B+2)}\right\}\right)
       \OPT_{\mathrm{open}}.
\]
\end{theorem}

\subsection{The algorithm and its certificates}
\label{subsec:uniform-certificates}

Our algorithm consists of running two parallel versions of Meyerson's algorithm, one that will give us the guarantee for singleton batches (pure random order) and one for nonsingleton ones. This allows us to use \Cref{lem:joint-n-minus-B-transfer}. Since the algorithm is the same, just run in parallel, we describe it once next. Namely, on the arrival of $v$, Meyerson's algorithm~\cite{meyerson2001online}
opens a facility at $v$ with probability
\[
    p_v=\min\left\{1,\frac{d(v,F_v)}{f}\right\},
\]
and otherwise it connects $v$ to a
closest open facility. The first client opens surely.
We write $F_v$ and $F_v^+$ for the facility sets immediately
before and after processing $v$. Similarly, $\mathcal H_v$
contains the execution history up to the arrival of $v$,
including its identity but not its lottery outcomes, while
$\mathcal H_v^+$ also includes the lotteries and decisions
made when processing $v$.

We use the same random variable for Meyerson's opening decision: that is, gor every client $v$, sample $U_v\sim\mathrm{Unif}(0,1)$, independently across clients and independently of both random arrival orders. We then realize Meyerson's decision by opening at $v$ exactly when $fU_v\le d(v,F_v)$. From here on, the main idea is to use a more restrictive event called a \emph{certified opening}, which follows the balanced-opening idea of Huang and Jiang~\cite{HuangJiang2026}.
In particular, fixed a comparison cluster $C$ with center $c$, we give each client $v\in C$ the key $Z_v=d(c,v)+fU_v$, and call
$v$ \emph{certified} if
\begin{equation}\label{eq:uniform-certificate}
 Z_v\le\min\{f,d(c,F_v)\}.
\end{equation}
Certification implies an opening at $v$, since
$fU_v\le d(c,F_v)-d(c,v)\le d(v,F_v)$, and hence $d(c,F_v^+)\le d(c,v)<Z_v$. 
The converse need not hold: there can be openings that are not certified. Hence, the analysis must charge the entire cost the algorithm incurs to these specific type of charges, which is exactly what we do in \Cref{lem:cost-to-certified}.
In addition, the reason the certification definition in \eqref{eq:uniform-certificate} has a cap at $f$ is because this will let us bound
the number of certifications directly. 
To this end, we define 
\[
 Y_v=f\mathbf1_{\{v\text{ is certified}\}},
 \qquad
 W_b=f\mathbf1_{\{\exists v\in C_b:\ v\text{ is certified}\}},
 \qquad W_C=\sum_bW_b.
\]
Hence, $Y_v$ counts the opening cost if client $v$ is certified and $W_b$ the opening cost if there is a client $v \in C_b$ that is certified. We also let $K_C=W_C/f$ be the number of batches containing a
certification from $C$. 

\subsection{Reducing cost to certified charges}
\label{subsec:uniform-cost-reduction}

\begin{lemma}\label{lem:cost-to-certified}
    It holds that
    \[
        \E[\ALG(C)]
 \le4\conn(C)+2\E\!\left[\sum_{v\in C}Y_v\right].
    \]
\end{lemma}
\begin{proof}
    The bound above follows by summing over $v\in C$ the inequality below and taking expectations:
    \begin{align*}
        &\E[\cost(v)\mid\mathcal H_v]\le2\min\{f,d(v,F_v)\} \le2\E[Y_v\mid\mathcal H_v]+4d(c,v).
    \end{align*}
    We start by proving the first inequality. Clearly, if $F_v=\varnothing$, the cost is $f$, whereas, if  $F_v\neq\varnothing$,
    \[
     \E[\cost(v)\mid\mathcal H_v]
     =fp_v+d(v,F_v)(1-p_v)
     \le2fp_v=2\min\{f,d(v,F_v)\}.
    \]
    Indeed, the connection term is zero if $p_v=1$, and is otherwise
    $d(v,F_v)=fp_v$. For the second inequality, observe that, conditional on the history $\mathcal H_v$ just before processing $v$, we have $\E[Y_v\mid\mathcal H_v]
 =[\min\{f,d(c,F_v)\}-d(c,v)]_+$. Therefore, by triangle inequality,
    \begin{align*}
     \min\{f,d(v,F_v)\}
     \le\min\{f,d(c,F_v)\}+d(c,v) &\le[\min\{f,d(c,F_v)\}-d(c,v)]_++2d(c,v)\\
     &=\E[Y_v\mid\mathcal H_v]+2d(c,v).
    \end{align*}
     To conclude, we note that for $F_v=\varnothing$, both capped distances are $f$, so the same
    calculation applies.
\end{proof}

\subsection{Summing charges within a batch}
\label{subsec:uniform-aggregation}

With a single opening price, the probability of certification
decreases with the distance to $c$. This lets us charge all
certifications after the first directly to connection distances.

\begin{lemma}[Total certified charge of a batch]
\label{lem:within-batch}
For every batch $b$,
\[
 \E\!\left[\sum_{v\in C_b}Y_v\right]
 \le\E[W_b]+\conn(C_b).
\]
\end{lemma}
\begin{proof}
We charge the first certification to $W_b$ and show that all
later certifications contribute at most $\conn(C_b)$ in
expectation. We fix a position in the shuffled batch and write $v$ for the client occupying it. Let $\mathcal H$ contain the execution
history before exposing $v$, and consider a history in which
no client of $C_b$ has yet been certified. We stress that $\mathcal H$ is the history before exposing the next
client $v$; revealing its identity gives $\mathcal H_v$, while
its lottery remains unexposed.
We next write $S\subseteq C_b$
for the unprocessed clients and $t=\min\{f,d(c,F)\}$ for the
current facility set $F$. We condition further on the event
$\mathcal E=\{v\in S,\ v\text{ is certified}\}$, whenever
$\Prb(\mathcal E\mid\mathcal H)>0$. Therefore, the first certified client's expected distance satisfies
\begin{align*}
    \E[d(c,v)\mid\mathcal H,\mathcal E] = \sum_{w\in S} d(c,w) \, \Prb(v=w\mid\mathcal H,\mathcal E) 
 =\frac{\sum_{w\in S}d(c,w)[t-d(c,w)]_+}
        {\sum_{w\in S}[t-d(c,w)]_+} \le\frac{\conn(S)}{|S|}.
\end{align*}
The second equality above follows since each remaining client is equally likely to come next, and since $w\in S$ is certified with probability $[t-d(c,w)]_+/f$. Thus,
\[
 \Prb(v=w\mid\mathcal H,\mathcal E)
 =\frac{[t-d(c,w)]_+}
        {\sum_{u\in S}[t-d(c,u)]_+}.
\]
For the inequality, we use the fact that these weights decrease with distance to $c$. Indeed, for every $u,w\in S$, $(d(c,u)-d(c,w)) ([t-d(c,u)]_+-[t-d(c,w)]_+)\le0$, and thus summing over $u,w\in S$ and rearranging gives
\[
 |S|\sum_{w\in S}d(c,w)[t-d(c,w)]_+
 \le\conn(S)\sum_{w\in S}[t-d(c,w)]_+.
\]
On event $\mathcal E$, a facility opens at $v$, and therefore every
later client $w\in S\setminus\{v\}$ has
$d(c,F_w)\le d(c,v)$. Its lottery is still fresh conditional
on $\mathcal H_w$, giving
\[
 \E[Y_w\mid\mathcal H_w]
 =[\min\{f,d(c,F_w)\}-d(c,w)]_+
 \le d(c,v).
\]
We sum over these later clients and take conditional
expectations again to obtain
\begin{align*}
 \E\!\left[\sum_{w\in S\setminus\{v\}}Y_w
       \;\middle|\;\mathcal H,\mathcal E\right]
 \le(|S|-1)\E[d(c,v)\mid\mathcal H,\mathcal E] \le\frac{|S|-1}{|S|}\conn(S)
 \le\conn(C_b).
\end{align*}
This bound holds at every possible position of the first
certification and for every preceding history. Averaging over
these mutually exclusive possibilities, and adding the first
charge $W_b$, proves the lemma.
\end{proof}

Summing over batches and using \Cref{lem:cost-to-certified}
immediately gives the following corollary.
\begin{corollary}\label{cor:uniform-first-reduction}
    It holds that
    \[
        \E[\ALG(C)]\le6\conn(C)+2f\,\E[K_C].
    \]
\end{corollary}

\subsection{Bounding the first certifications}
\label{subsec:uniform-records}

For each batch with $C_b\neq\varnothing$, let
$Q_b=\min_{v\in C_b}Z_v$; batches with $C_b=\varnothing$
contribute zero and can be omitted.
We give two bounds on $K_C$: one in terms of the number of
batches, and one in terms of the number of clients and their
connection distances.

\begin{lemma}[Record bound]
\label{lem:records}
For every comparison cluster $C$, $\E[K_C]\le H_B$.
\end{lemma}
\begin{proof}
After processing any $v\in C$,
$\min\{f,d(c,F_v^+)\}<Z_v$.
Indeed, if the threshold was already below $Z_v$, it remains
below; otherwise $v$ is certified and
$\min\{f,d(c,F_v^+)\}\le d(c,v)<Z_v$.
Applying this to a client attaining $Q_{b'}$ shows that the
threshold remains below $Q_{b'}$ after batch $b'$.
Thus a certified client $v$ in a later batch $b$ gives
\[
 Q_b\le Z_v\le\min\{f,d(c,F_v)\}<Q_{b'}.
\]
Every batch counted by $K_C$ is therefore a lower record of
the batch keys. Write $Q_1,\ldots,Q_s$ for the keys of the
batches meeting $C$, where $s\le B$, and let
$\pi_1,\ldots,\pi_s$ be their arrival order.
Conditional on these keys, which are distinct almost surely,
$\pi$ remains a uniformly random permutation. Hence
\begin{align*}
 \E[K_C\mid Q_1,\ldots,Q_s]
 \le\sum_{r=1}^{s}
   \Prb\!\left(
      Q_{\pi_r}=\min_{\ell\le r}Q_{\pi_\ell}
      \;\middle|\;Q_1,\ldots,Q_s\right)=\sum_{r=1}^{s}\frac1r
 =H_s\le H_B.
\end{align*}
The equality follows because each of the first $r$ positions
is equally likely to contain their smallest key.
Taking expectations proves the lemma.
\end{proof}

\begin{lemma}[Progress and records]
\label{lem:records-II}
For every $\rho>1$,
\[
    \E[K_C]\le3+\frac{\log n}{\log\rho}
       +\log\left(1+\frac{(\rho-1)\conn(C)}{f}\right).
\]
\end{lemma}
\begin{proof}
Count the first certified client of $C_b$ in each batch.
Their number is exactly $K_C$. The certification threshold
$\min\{f,d(c,F)\}$ starts at most $f$ and never increases.
We distinguish three cases.

\noindent\emph{(1) Multiplicative progress.}
If $\min\{f,d(c,F_v)\}>f/n$ and
$d(c,v)\le\min\{f,d(c,F_v)\}/\rho$, certification gives
\[
 \min\{f,d(c,F_v^+)\}\le d(c,v)
 \le\frac{\min\{f,d(c,F_v)\}}{\rho}.
\]
Before the $r$th such certification, $f/n < \min\{f,d(c,F_v)\} \le f/\rho^{r-1}$, and thus $r \le 1+\log n/\log\rho$, which means that this case contributes at most $1+\log n/\log\rho$.\medskip

\noindent\emph{(2) Small certification threshold.}
If $\min\{f,d(c,F_v)\}\le f/n$, then
\[
 \Prb(v\text{ is certified}\mid\mathcal H_v)
 =\frac{[\min\{f,d(c,F_v)\}-d(c,v)]_+}{f}
 \le\frac1n.
\]
For every history, a client either fails the small-threshold
condition and contributes zero to this case, or is certified
with probability at most $1/n$. The expected count is therefore
at most $\sum_{v\in C}1/n\le1$.\medskip

\noindent\emph{(3) Remaining certifications.}
Here $d(c,v)>\min\{f,d(c,F_v)\}/\rho$, so
\[
 fU_v\le\min\{f,d(c,F_v)\}-d(c,v)<(\rho-1)d(c,v).
\]
Call a batch \emph{suspect} if it contains a client $v\in C_b$ satisfying
$fU_v<(\rho-1)d(c,v)$, and let $N$ count these batches.
A union bound gives
\[
 \E[N]\le\sum_{v\in C}\Prb(fU_v<(\rho-1)d(c,v))
 =\sum_{v\in C}\min\left\{1,\frac{(\rho-1)d(c,v)}f\right\}
 \le\frac{(\rho-1)\conn(C)}f.
\]
Conditioning on all lottery variables fixes the suspect batches
and their keys, while their relative order remains uniform.
Every remaining counted batch is a lower record also among
suspect batches, by the preceding lemma. Its conditional
expected count is therefore at most $\sum_{r=1}^N1/r=H_N$.
Taking expectations and applying Jensen's inequality,
\[
 \E[H_N]\le1+\E[\log(1+N)]
 \le1+\log(1+\E[N])
 \le1+\log\left(1+\frac{(\rho-1)\conn(C)}f\right).
\]
Adding the three bounds proves the lemma.
\end{proof}

We now combine all of the above to finish the proof of \Cref{thm:uniform-batched}.

\begin{proof}[Proof of \Cref{thm:uniform-batched}]
First, let us choose $\rho=\exp(\max\{1,\sqrt{\log(n+1)}\})$, and apply \Cref{cor:uniform-first-reduction,lem:records,lem:records-II}, to obtain
\begin{align*}
 \E[\ALG(C)]
 &\le 8\conn(C)+12f \cdot \min\left\{\log(B),\sqrt{\log(n)}\right\}.
\end{align*}
Then, summing over the optimal clusters and applying \Cref{lem:joint-n-minus-B-transfer} proves
\[
 \E[\ALG_{\mathrm{combined}}]\le O(1)\OPT_{\mathrm{conn}}
 +O\!\left(\min\left\{\log(B+1),\sqrt{\log(n-B+2)}\right\}\right)
       \OPT_{\mathrm{open}}. \qedhere
\]
\end{proof}

\section{Nonuniform opening costs}
\label{sec:nonuniform-costs}
\label{sec:nonuniform}

We extend the uniform analysis using charges at powers of two: we compare clients’ total charges across levels within each batch, then use geometric charging to apply the uniform record and progress bounds.

\begin{theorem}[Nonuniform opening costs]
\label{thm:nonuniform-batched}\label{nonunif:main}
For $n\ge1$ clients in $1\le B\le n$ batches, there exists an algorithm satisfying
\[
    \E[\ALG]\le O(1)\OPT_{\mathrm{conn}}
 +O\!\left(\min\left\{\log(B+1),\sqrt{\log(n-B+2)}\right\}\right)
       \OPT_{\mathrm{open}}.
\]
\end{theorem}

\subsection{The algorithm and its certificates}

Before describing the algorithm and carrying out its analysis, we simplify our problem by rounding up opening costs up to powers of two, so
$f(y)\le\widehat f(y)<2f(y)$, and letting $f_1<\cdots<f_m$ be the
distinct rounded costs. We also denote the set of facilities of cost at most $f_j$ as
$\mathsf M_j=\{y:\widehat f(y)\le f_j\}$, with
$d_v^{(j)}=d(v,\mathsf M_j)$ and $d_v^{(0)}=\infty$.
Every client location is permitted, so $d_v^{(m)}=0$.

We use again Meyerson's marginal rule~\cite{meyerson2001online}, which uses the facility
set $F_v$ present before processing $v$. More specifically, let us write
$\Delta_v^{(0)}=d(v,F_v)$ and
$\Delta_v^{(j)}=\min\{d(v,F_v),d_v^{(j)}\}$, and, independently at each level $j$, we let the algorithm select a closest point of
$F_v\cup\mathsf M_j$ with probability
\begin{equation}\label{eq:expanded-marginal-probability}
 p_v^{(j)}=\min\left\{1,
          \frac{\Delta_v^{(j-1)}-\Delta_v^{(j)}}{f_j}\right\},
\end{equation}
opening every selected facility that is not already open, paying its
original cost, and connecting $v$ to a closest facility in the resulting
set $F_v^+$.  All lotteries use the same pre-client set $F_v$ and fixed
distance tie breaking, and when $F_v=\varnothing$, the first lottery
succeeds surely. We realize that the right-hand side of \eqref{eq:expanded-marginal-probability} is equal to the marginal decrease of the distance to the closest facility when the set of facilities $\mathsf M_j\setminus \mathsf M_{j-1}$ of level $j$ is added to $\mathsf M_{j-1} \cup F_v$. 

We aim to mirror the coupling used in the uniform case. To this end, using independent $U_v^{(j)}\sim\mathrm{Unif}(0,1)$ (also independent
of all arrival orders), the level-$j$ lottery succeeds precisely when
\begin{equation}\label{eq:expanded-static-lottery}
 d_v^{(j)}+f_jU_v^{(j)}
 \le\min\{d(v,F_v),d_v^{(j-1)}\}.
\end{equation}
Indeed, its probability is
$\min\{1,[\min\{d(v,F_v),d_v^{(j-1)}\}-d_v^{(j)}]_+/f_j\}
=p_v^{(j)}$. Upon success, positivity of $U_v^{(j)}$ implies that
the selected facility belongs to $\mathsf M_j$ and is at distance
$d_v^{(j)}$ from $v$.

We next fix a comparison cluster $C$ with center $c$, and write $f_C=f_{j_C}$
for its rounded opening cost. Similarly to the uniform case, for $v \in C$, we give each client-level pair $(v,j)$ the static key

\begin{equation}\label{def:client-level-static-key}
Z_v^{(j)}=d(c,v)+d_v^{(j)}+f_jU_v^{(j)},
\end{equation} setting it to $\infty$
if $d_v^{(j)}+f_jU_v^{(j)}>d_v^{(j-1)}$.
We say that the client-level pair $(v,j)$ is \emph{certified} if its key is finite and
$Z_v^{(j)}\le d(c,F_v)$. Certification implies a successful lottery,
since $d(v,F_v)\ge d(c,F_v)-d(c,v)$, and gives
\begin{equation}\label{eq:expanded-certified-progress}
 d(c,F_v^+)\le d(c,v)+d_v^{(j)}<Z_v^{(j)}.
\end{equation}
We stress that being certified is a sufficient condition for opening but is not necessary. Following the above, we define
\[
    Y_v=\sum_{j\le j_C}f_j\mathbf1_{\{(v,j)\text{ is certified}\}}
\]
as the total cost of certified openings induced by client $v$ within cost level $j_C$ (the rounded opening cost of the optimal facility serving cluster $C$). As we will see, we retain only up until this level because all the more expensive levels can be charged directly to the optimal connection cost.
Similarly, for each batch $C_b=C\cap b$, let $W_b$ charge $f_j$ once for each level
$j\le j_C$ with at least one certified client in $C_b$, i.e.,
\[
    W_b:=\sum_{j \le j_C} f_j\, \mathbf 1_{\{\exists\,v\in C_b:\ (v,j)\text{ is certified}\}},
\]
and put $W_C=\sum_bW_b$. Thus \(W_b\) pays \(f_j\) once if at least one client of \(C_b\) is certified at level \(j\), regardless of how many such certifications occur, while $Y_v$ counts all certified charges.

\subsection{Reducing cost to certified charges}

\begin{lemma}[Reduction to certified charges]
\label{lem:nonuniform-certified-cost}
For every comparison cluster $C$,
\[
 \E[\ALG(C)]\le6\conn(C)+2\E\left[\sum_{v\in C}Y_v\right].
\]
\end{lemma}

Our strategy to prove the above lemma will be similar to the uniform case. We will in fact prove that
\[
 \E[\cost(v)\mid\mathcal H_v]
 \le6d(c,v)+2\E[Y_v\mid\mathcal H_v],
\]
for every client.
Once this is shown, the lemma simply follows by taking expectations and summing.
To this end, let us first fix $v$ and condition on $\mathcal H_v$, proving the following claim.
\begin{claim}\label{cl:nonuniform-lottery-cost}
    Fix $v$ and condition on $\mathcal H_v$. Then, it holds that
    \begin{equation}
 \E[\cost(v)\mid\mathcal H_v]
 \le\sum_{j=1}^m f_jp_v^{(j)}
      +\E[d(v,F_v^+)\mid\mathcal H_v]
 \le2\sum_{j=1}^m f_jp_v^{(j)}.
\end{equation}
\end{claim}
\begin{proof}
    Fixed $v$ and conditioned on $\mathcal H_v$, we know that each newly opened facility
is selected by a successful lottery, and a level-$j$ lottery selects
at most one facility, whose opening cost is at most $f_j$. Hence,
for every outcome,
\[
 \sum_{y\in F_v^+\setminus F_v}f(y)
 \le\sum_{j=1}^m f_j\,
       \mathbf1_{\{\text{level-$j$ lottery succeeds}\}}.
\] 
Thus, taking conditional expectations gives
\begin{align*}
 \E\!\left[\sum_{y\in F_v^+\setminus F_v}f(y)
            \;\Big|\;\mathcal H_v\right]
 \le\sum_{j=1}^m f_j\,
       \Prb(\text{level-$j$ lottery succeeds}\mid\mathcal H_v) =\sum_{j=1}^m f_jp_v^{(j)}.
\end{align*}
We next show that $\sum_{j=1}^m f_jp_v^{(j)}$ also bounds the expected
connection cost. The key observation is that the final connection
distance is at most the sum of the marginal gaps at failed levels. Indeed, suppose first that $F_v\ne\varnothing$ and at least one lottery
succeeds, and let $k$ be the largest successful level. Its selected
facility is within distance $\Delta_v^{(k)}$ of $v$, and every
level above $k$ fails. Since $\Delta_v^{(m)}=0$, we obtain
\[
 d(v,F_v^+)
 \le\Delta_v^{(k)}
 =\sum_{j>k}\left(\Delta_v^{(j-1)}-\Delta_v^{(j)}\right)
 \le\sum_{j=1}^m\left(\Delta_v^{(j-1)}-\Delta_v^{(j)}\right)
       \mathbf1_{\{\text{level-$j$ lottery fails}\}},
\]
where the last inequality adds
the nonnegative gaps of any failed levels below $k$.
If every lottery fails, the last sum telescopes to
$\Delta_v^{(0)}=d(v,F_v)$, so the same bound holds.
Taking conditional expectations, when $F_v\ne\varnothing$ we obtain
\[
 \E[d(v,F_v^+)\mid\mathcal H_v]
 \le\sum_{j=1}^m
   \left(\Delta_v^{(j-1)}-\Delta_v^{(j)}\right)(1-p_v^{(j)})
 \le\sum_{j=1}^m f_jp_v^{(j)}.
\]
The last inequality holds term by term: if $p_v^{(j)}=1$,
the contribution is zero, and otherwise,
$\Delta_v^{(j-1)}-\Delta_v^{(j)}=f_jp_v^{(j)}$ and
$1-p_v^{(j)}\le1$.
If $F_v=\varnothing$, the first lottery succeeds surely,
so the same telescoping argument and expectation bound apply
with both sums starting at $j=2$, where all gaps are finite.
Adding the nonnegative term $f_1p_v^{(1)}$ to the final bound
gives the same conclusion.
Combining this connection-cost bound with the opening-cost
bound proved above yields, in all cases,
\[
    \E[\cost(v)\mid\mathcal H_v]
 \le\sum_{j=1}^m f_jp_v^{(j)}
      +\E[d(v,F_v^+)\mid\mathcal H_v]
 \le2\sum_{j=1}^m f_jp_v^{(j)}. \qedhere
\]
\end{proof}

It remains to compare the lottery charges to the expected certified cost $Y_v$, as we do next.
\begin{claim}\label{cl:lottery-to-certified}
    Fix $v$ and condition on $\mathcal H_v$. Then, it holds that
    \[
        \sum_{j=1}^m f_jp_v^{(j)} \le \E[Y_v\mid\mathcal H_v]+3d(c,v).
    \]
\end{claim}
\begin{proof}
We will show that replacing lottery charges by certified charges
at all levels loses at most $2d(c,v)$ in expectation. Removing
the levels above $j_C$ then loses at most another $d(c,v)$,
giving the desired bound. 

Condition on $\mathcal H_v$ and suppose first that
$F_v\ne\varnothing$. The triangle inequality gives
\[
 [d(c,F_v)-d(c,v)]_+
 \le d(v,F_v)
 \le [d(c,F_v)-d(c,v)]_++2d(c,v).
\]
At level $j$, the expected lottery charge is
$\min\{f_j,\Delta_v^{(j-1)}-\Delta_v^{(j)}\}$, whereas the
expected certified charge is
\begin{align*}
    &f_j\Prb((v,j)\text{ is certified}\mid\mathcal H_v)\\
 =~
 &\min\left\{f_j,\,
 \min\{[d(c,F_v)-d(c,v)]_+,d_v^{(j-1)}\}
 -
 \min\{[d(c,F_v)-d(c,v)]_+,d_v^{(j)}\}
 \right\}.
\end{align*}
The marginal length in the second expression is no larger
than $\Delta_v^{(j-1)}-\Delta_v^{(j)}$, since its cutoff is
no larger than $d(v,F_v)$. Using
$\min\{f_j,x\}-\min\{f_j,y\}\le x-y$ for $x\ge y\ge0$,
we obtain
\begin{align*}
 &\sum_{j=1}^m f_jp_v^{(j)}
 -\E\!\left[\sum_{j=1}^m f_j
   \mathbf1_{\{(v,j)\text{ is certified}\}}
   \;\middle|\;\mathcal H_v\right]\\
 \le \,
 &\sum_{j=1}^m\left(
 \Delta_v^{(j-1)}-\Delta_v^{(j)}
 -\min\{[d(c,F_v)-d(c,v)]_+,d_v^{(j-1)}\}
 +\min\{[d(c,F_v)-d(c,v)]_+,d_v^{(j)}\}
 \right)\\
 =\, &d(v,F_v)-[d(c,F_v)-d(c,v)]_+
 \le2d(c,v).
\end{align*}
The equality follows by telescoping, using
$\Delta_v^{(0)}=d(v,F_v)$, $d_v^{(0)}=\infty$, and
$\Delta_v^{(m)}=d_v^{(m)}=0$.
If $F_v=\varnothing$, success and certification both reduce to
$d_v^{(j)}+f_jU_v^{(j)}\le d_v^{(j-1)}$. their expected charges are therefore equal, so the bound
also holds.

It remains to account for the certified charges above level
$j_C$, which are not included in $Y_v$: each is bounded in
expectation by the length of its marginal interval, so
\[
 \E\!\left[\sum_{j>j_C}f_j
      \mathbf1_{\{(v,j)\text{ is certified}\}}
      \;\middle|\;\mathcal H_v\right]
 \le\sum_{j>j_C}(d_v^{(j-1)}-d_v^{(j)})
 =d_v^{(j_C)}
 \le d(c,v).
\]
The equality uses $d_v^{(m)}=0$, and the last inequality follows
from $c\in\mathsf M_{j_C}$. Substituting this bound above and
using the definition of $Y_v$ gives
\[
 \sum_{j=1}^m f_jp_v^{(j)}
 \le \E[Y_v\mid\mathcal H_v]+3d(c,v). \qedhere
\]
\end{proof}

We can now finish the proof of \Cref{lem:nonuniform-certified-cost}.

\begin{proof}
Combining the bound in \Cref{cl:nonuniform-lottery-cost} with the one in \Cref{cl:lottery-to-certified} reaches the desired final bound:
\begin{align*}
 \E[\cost(v)\mid\mathcal H_v]
 &\le2\sum_{j=1}^m f_jp_v^{(j)} \le4d(c,v)+2\E\!\left[
       \sum_{j=1}^m f_j\mathbf1_{\{(v,j)\text{ is certified}\}}
       \;\middle|\;\mathcal H_v\right]\\
 &\le6d(c,v)+2\E[Y_v\mid\mathcal H_v].
\end{align*}
Taking expectations and summing over $v\in C$ proves the lemma.
\end{proof}

\subsection{Summing charges within a batch}
\label{subsec:expanded-nonuniform-aggregation}

As we saw in the earlier section, \Cref{lem:nonuniform-certified-cost} bounds each client's cost
in terms of its certified charge and its distance to $c$.
We next bound, as in \Cref{lem:within-batch} for the uniform opening cost case, the total certified
charge of a batch by its connection cost and the charges for
first certifications. Here, $W_b$ charges $f_j$ once for each
level $j\le j_C$ at which some client in $C_b$ is certified.

We first bound the charges of the half of the clients closest
to $c$, then pair each remaining client with a distinct client
in this nearer half and compare their total certified charges.
Unlike in the uniform case, being closer to $c$ need not give
a higher certification probability at a particular level:
a nearer client may have zero marginal gap at that level,
even when a farther client's gap is positive.

\begin{lemma}[Total certified charge of a batch]
\label{lem:nonunif-aggregation}
For every batch $b$,
\[
 \E\left[\sum_{v\in C_b}Y_v\right]
 \le6\E[W_b]+14\conn(C_b).
\]
\end{lemma}

To prove the above lemma, we may assume that $C_b\neq\emptyset$, since the claim is trivial
when $C_b=\emptyset$. Let us call the
$\lceil|C_b|/2\rceil$ clients closest to $c$ \emph{near}, breaking
ties in a fixed way, and the others \emph{far}. We first prove the following claim about near clients, and then include far ones.

\begin{claim}[Near clients]\label{cl:near-clients}
    It holds that
    \[
        \E\left[\sum_{v\,\mathrm{near}}Y_v\right]
 \le\E[W_b]+2\conn(C_b).
 \]
\end{claim}

\begin{proof}

We first observe that the farthest near client and all far clients give
$\lfloor|C_b|/2\rfloor+1\ge\lceil|C_b|/2\rceil$ clients at least
as far from $c$ as every near client $v$, and therefore, 
\[
 \conn(C_b)\ge
 \left(\left\lfloor\frac{|C_b|}{2}\right\rfloor+1\right)d(c,v)
 \ge \left\lceil\frac{|C_b|}{2}\right\rceil\,d(c,v).
\]
In particular, $d(c,v)\le\conn(C_b)/\lceil|C_b|/2\rceil$.
We will charge at most $W_b$ for one certification per level,
and bound the remaining expected charge by
$2\conn(C_b)/\lceil|C_b|/2\rceil$ per near client.
For each level $j\le j_C$, consider the condition
\begin{equation}\label{eq:nonuniform-near-condition}
 d(c,F)\le d(c,\mathsf M_j)
                 +\frac{2\conn(C_b)}{\lceil|C_b|/2\rceil},
\end{equation}
tested before the current client's lotteries. A certification
of a near client $v$ makes this condition true, because
\[
 d(c,F_v^+)\le d(c,v)+d_v^{(j)}
 \le2d(c,v)+d(c,\mathsf M_j)
 \le\frac{2\conn(C_b)}{\lceil|C_b|/2\rceil}
      +d(c,\mathsf M_j).
\]
Once true, it remains true, since facilities are only added.
Thus, at each level, at most one near client can be certified
while the condition is false. Such a certification costs $f_j$,
which is included in $W_b$, and their total charge is therefore
at most $W_b$ for every execution.

We next bound the charges incurred when the condition is already
true. Before a near client $v$ is processed, let $k=k(v)\le j_C$
be the highest level satisfying \eqref{eq:nonuniform-near-condition},
or $k=0$ if there is none. Note that this index is determined by
$\mathcal H_v$. The levels satisfying the condition form a
prefix $1,\ldots,k$, since
$d(c,\mathsf M_j)\ge d(c,\mathsf M_k)$ for $j\le k$, and so, for a nonempty prefix, the conditional expected charge satisfies
\begin{align*}
 \E\!\left[\sum_{j\le k}f_j
       \mathbf1_{\{(v,j)\text{ is certified}\}}
       \;\middle|\;\mathcal H_v\right] &=\sum_{j\le k}\min\left\{f_j,
   \left[\min\{[d(c,F_v)-d(c,v)]_+,d_v^{(j-1)}\}
         -d_v^{(j)}\right]_+\right\}\\
 &\le[d(c,F_v)-d(c,v)-d_v^{(k)}]_+\\
 &\le[d(c,F_v)-d(c,\mathsf M_k)]_+
 \le\frac{2\conn(C_b)}{\lceil|C_b|/2\rceil}.
\end{align*}
For the first inequality, the summands are lengths of disjoint
portions of the intervals
$(d_v^{(j)},\min\{d_v^{(j)}+f_j,d_v^{(j-1)}\}]$. For $j\le k$, these portions
all lie above $d_v^{(k)}$ and below
$[d(c,F_v)-d(c,v)]_+$, so their total length is at most
$[d(c,F_v)-d(c,v)-d_v^{(k)}]_+$.
The second inequality uses
$d(c,\mathsf M_k)\le d(c,v)+d_v^{(k)}$,
and the last uses \eqref{eq:nonuniform-near-condition}.
An empty prefix contributes zero, including when $F_v=\emptyset$.
Taking expectations and summing over the near clients now gives
\[
 \E\left[\sum_{v\text{ near}}Y_v\right]
 \le\E[W_b]
   +\left\lceil\frac{|C_b|}{2}\right\rceil
      \frac{2\conn(C_b)}{\lceil|C_b|/2\rceil}
 =\E[W_b]+2\conn(C_b). \qedhere
\]
\end{proof}

We now turn to the far clients, by proving an inequality that allows us to pair them with a near client at a small cost.

\begin{claim}[Far clients]
\label{cl:far-clients}
For distinct $u,v\in C_b$,
\[
 \E[Y_u+Y_v]\le6\E[Y_u]+2d(u,v).
\]
\end{claim}
\begin{proof}
Let us condition on the history $\mathcal H$ immediately before either client
arrives, and let $F$ be the current facility set. Write $u\prec v$ if $u$
arrives first. Swapping their positions preserves $\mathcal H$, so each
order has conditional probability $1/2$, and both clients' lotteries
remain fresh. We first prove
\begin{equation}\label{eq:nonuniform-first-comparison}
 \E[Y_v\mid\mathcal H,v\prec u]
 \le2\E[Y_u\mid\mathcal H,u\prec v]+2d(u,v).
\end{equation}
Let $\ell\le j_C$ be the largest level contributing its full
expected charge $f_\ell$ when $u$ arrives first; equivalently,
$\Prb((u,\ell)\text{ is certified}\mid\mathcal H,u\prec v)=1$.
Set $\ell=0$ and $f_0=0$ if there is no such level.
For $w\in\{u,v\}$ arriving first, the expected charge above $\ell$ satisfies
\begin{align*}
 &\sum_{j=\ell+1}^{j_C}\min\left\{f_j,
   \left[\min\{[d(c,F)-d(c,w)]_+,d_w^{(j-1)}\}
         -d_w^{(j)}\right]_+\right\}\\
 \le
 &\min\{[d(c,F)-d(c,w)]_+,d_w^{(\ell)}\}
 -\min\{[d(c,F)-d(c,w)]_+,d_w^{(j_C)}\},
\end{align*}
where the inequality uses $\min\{f_j,x\}\le x$ and telescoping. Note that it is an
equality for $u$, since no summand above $\ell$ reaches its cap.
For $F\neq\varnothing$, the triangle inequality gives
$|d(c,u)-d(c,v)|\le d(u,v)$ and
$|d_u^{(j)}-d_v^{(j)}|\le d(u,v)$ for $j\ge1$.
Positive parts preserve this bound, and
$|\min\{a,b\}-\min\{a',b'\}|
\le\max\{|a-a'|,|b-b'|\}$.
Thus each endpoint changes by at most $d(u,v)$.
The levels through $\ell$ cost $v$ at most
$\sum_{j\le\ell}f_j\le2f_\ell$, since the prices are distinct powers
of two, whereas $u$ already pays $f_\ell$ at level $\ell$.
Consequently,
\begin{align*}
 \E[Y_v\mid\mathcal H,v\prec u]
 &\le2f_\ell+\left(\E[Y_u\mid\mathcal H,u\prec v]-f_\ell\right)+2d(u,v)\\
 &\le2\E[Y_u\mid\mathcal H,u\prec v]+2d(u,v),
\end{align*}
as claimed, since
$f_\ell\le\E[Y_u\mid\mathcal H,u\prec v]$.
For $\ell=0$, the prefix is empty and the first endpoint is
$[d(c,F)-d(c,w)]_+$, which also changes by at most $d(u,v)$.
If $F=\varnothing$, level $1$ contributes $f_1$, so $\ell\ge1$.
The endpoints are then $d_w^{(\ell)}$ and $d_w^{(j_C)}$, and the
same triangle-inequality bound applies.

Finally, arriving later can only decrease a client's expected charge:
facilities are only added and its lotteries remain fresh. Averaging
over the two equally likely orders therefore gives
\begin{align*}
 \E[Y_u+Y_v\mid\mathcal H]
 \le\E[Y_u\mid\mathcal H,u\prec v]
      +\E[Y_v\mid\mathcal H,v\prec u] &\le3\E[Y_u\mid\mathcal H,u\prec v]+2d(u,v)\\
 &\le6\E[Y_u\mid\mathcal H]+2d(u,v).
\end{align*}
For the first inequality, in each order the later client's
expected charge is at most its first-arrival expectation.
The last inequality uses
$\E[Y_u\mid\mathcal H]\ge 1/2 \cdot \E[Y_u\mid\mathcal H,u\prec v]$.
Taking expectations proves the claim.
\end{proof}

Finally, we can conclude the proof of \Cref{lem:nonunif-aggregation}.

\begin{proof}[Proof of \Cref{lem:nonunif-aggregation}]

We pair each far client $v$ with a distinct near client $u$,
fixing the pairing independently of the arrival order and lotteries.
By \Cref{cl:far-clients}, we have that
$\E[Y_u+Y_v]\le6\E[Y_u]+2d(u,v)$ for every pair. Hence
\begin{align*}
 \E\left[\sum_{v\in C_b}Y_v\right]
 \le6\E\left[\sum_{u\text{ near}}Y_u\right]
      +2\sum_{(u,v)\text{ paired}}d(u,v)
 \le 6\E[W_b]+14\conn(C_b).
\end{align*}
The first inequality also covers the possible unpaired near client,
since its charge is nonnegative. The second uses
\Cref{cl:near-clients} and the triangle inequality:
$\sum_{(u,v)\text{ paired}}d(u,v)
\le\sum_{(u,v)\text{ paired}}(d(c,u)+d(c,v))
\le\conn(C_b)$, because the pairs are disjoint.
\end{proof}

Summing \Cref{lem:nonunif-aggregation} over the batches and
applying \Cref{lem:nonuniform-certified-cost} gives the following
proposition.

\begin{proposition}[Reduction to first certificates]
\label{prop:nonunif-reduction}
For every cluster $C$,
\[
    \E[\ALG(C)]\le34\conn(C)+12\E[W_C].
\]
\end{proposition}

\subsection{Bounds at each opening-cost level}
\label{subsec:expanded-nonuniform-records}

We now use a similar analysis of first certifications cost as the uniform case, proceeding level by level. Indeed, let $K_C^{(j)}$ count batches
containing a level-$j$ certification from $C$, and let $L_C^{(j)}$
count those containing a \emph{regular} level-$j$ certification, which is called such if $Z_v^{(j)}\le d(c,\mathsf M_j)+f_j$.

\begin{proposition}[Bounds at each level]
\label{prop:short-nonuniform-records}\label{nonunif:records}
\label{nonunif:level-lemma}
For every $j\le j_C$ and $\rho>1$,
\begin{align}
 \E[K_C^{(j)}]&\le H_B,
 \label{eq:nonuniform-level-harmonic}\\
 \E[L_C^{(j)}]&\le3+\frac{\log n}{\log\rho}
       +\log\left(1+\frac{2(\rho-1)\conn(C)}{f_j}\right).
 \label{eq:short-level-bound}
\end{align}
\end{proposition}

\noindent In the remainder of the argument, let us recall that $j\le j_C$ and let us fix the following notation. Namely, subtract the baseline $d(c,\mathsf M_j)$ from the
current distance and the finite client keys, defining
\[
 q_j(F)=[d(c,F)-d(c,\mathsf M_j)]_+,
 \qquad
 \xi_v^{(j)}=d(c,v)+d_v^{(j)}-d(c,\mathsf M_j),
 \qquad
 Q_b^{(j)}=\min_{v\in C_b}\{Z_v^{(j)}-d(c,\mathsf M_j)\}.
\]
Thus $q_j(F)$ measures the excess distance to $c$, which a certification reduces to at most $\xi_v^{(j)}$ as we will show. Similarly, $Q_b^{(j)}$
is the minimum shifted key in batch $b$. It is understood that $q_j(\varnothing)=\infty$ and that discarded keys remain infinite.

\begin{claim}[Certification and progress]
\label{cl:nonuniform-level-correspondence}
For every $v\in C$, $0\le\xi_v^{(j)}\le2d(c,v)$. Moreover,
almost surely,
\begin{align}
 (v,j)\text{ is certified}
 &\Longleftrightarrow Z_v^{(j)}<\infty\ \text{ and }\
      \xi_v^{(j)}+f_jU_v^{(j)}\le q_j(F_v),
 \label{eq:expanded-residual-certificate}\\
 q_j(F_v^+)&\le\xi_v^{(j)}
       \quad\text{whenever $(v,j)$ is certified}.
 \label{eq:expanded-residual-progress}
\end{align}
For regular certification, replace $q_j(F_v)$ in
\eqref{eq:expanded-residual-certificate} by $\min\{f_j,q_j(F_v)\}$.
\end{claim}
\begin{proof}
The triangle inequality gives
$d(c,\mathsf M_j)-d(c,v)\le d_v^{(j)}
\le d(c,\mathsf M_j)+d(c,v)$.
Adding $d(c,v)-d(c,\mathsf M_j)$ proves
$0\le\xi_v^{(j)}\le2d(c,v)$.
For a finite key,
\[
 Z_v^{(j)}-d(c,\mathsf M_j)
 =\xi_v^{(j)}+f_jU_v^{(j)}>0
 \qquad\text{almost surely}.
\]
Subtracting the baseline from the certification test
$Z_v^{(j)}\le d(c,F_v)$ therefore gives
\eqref{eq:expanded-residual-certificate}; the positive part does
not change this comparison with a positive key.
Regularity additionally requires this shifted key to be at most
$f_j$, giving the capped rule.
Finally, a certification ensures an open facility within
$d_v^{(j)}$ of $v$, so
\[
 q_j(F_v^+)
 \le[d(c,v)+d_v^{(j)}-d(c,\mathsf M_j)]_+
 =[\xi_v^{(j)}]_+=\xi_v^{(j)}. \qedhere
\]
\end{proof}

\begin{claim}[Record bound]
\label{cl:nonuniform-level-records}
One has $\E[K_C^{(j)}]\le H_B$.
\end{claim}
\begin{proof}
The claim follows by a simple record argument: we show that, for a fixed level $j$, a batch key is smaller than the key of every earlier batch. 

First, consider a client $v$ whose level-$j$ key is finite.
We show that, after processing $v$, the excess distance
$q_j(F_v^+)$ is smaller than its shifted key
$\xi_v^{(j)}+f_jU_v^{(j)}$.
If $q_j(F_v)$ was already smaller than this key, opening
facilities can only decrease it, so
$q_j(F_v^+)\le q_j(F_v)<\xi_v^{(j)}+f_jU_v^{(j)}$.
Otherwise, the certification rule certifies $(v,j)$, and the
resulting opening gives
$q_j(F_v^+)\le\xi_v^{(j)}<\xi_v^{(j)}+f_jU_v^{(j)}$ by \cref{cl:nonuniform-level-correspondence}.

Now consider a batch $b'$ with $Q_{b'}^{(j)}<\infty$, and apply
this observation to a client attaining its minimum shifted key
$Q_{b'}^{(j)}$. After that client is processed, $q_j(F)$ is
strictly below $Q_{b'}^{(j)}$ and remains so, since facilities
are only added. Thus, if a later batch $b$ contains a certified
client $v$, then
\[
 Q_b^{(j)}
 \le\xi_v^{(j)}+f_jU_v^{(j)}
 \le q_j(F_v)<Q_{b'}^{(j)}.
\]
The first inequality uses the definition of the batch minimum,
the second uses certification, and the last follows from the
preceding observation. Hence a batch containing a certification
has a smaller key than every earlier batch.

Every batch counted by $K_C^{(j)}$ is therefore a lower record
of the batch keys. Write $Q_1^{(j)},\ldots,Q_s^{(j)}$ for the
finite batch keys, indexed independently of their arrival order,
where $s\le B$, and let $\pi_1,\ldots,\pi_s$ be their arrival order.
Conditional on $s$ and these keys, which are distinct almost surely,
$\pi$ remains a uniformly random permutation. Hence
\begin{align*}
 \E[K_C^{(j)}\mid Q_1^{(j)},\ldots,Q_s^{(j)}]
 \le\sum_{r=1}^{s}
   \Prb\!\left(
      Q_{\pi_r}^{(j)}=\min_{\ell\le r}Q_{\pi_\ell}^{(j)}
      \;\middle|\;Q_1^{(j)},\ldots,Q_s^{(j)}\right) =\sum_{r=1}^{s}\frac1r=H_s\le H_B.
\end{align*}
Taking expectations gives $\E[K_C^{(j)}]\le H_B$.
\end{proof}

\begin{claim}[Bound for regular certifications]
\label{cl:nonuniform-level-regular}
For every $\rho>1$,
\[
 \E[L_C^{(j)}]\le3+\frac{\log n}{\log\rho}
 +\log\left(1+
       \frac{(\rho-1)\sum_{v\in C}\xi_v^{(j)}}{f_j}\right).
\]
\end{claim}
\begin{proof}
We need to bound the expected number of batches containing a
regular level-$j$ certification from $C$. In each such batch,
we count only the first one, so that the number counted is exactly $L_C^{(j)}$. We also recall that a regular certification requires
$\xi_v^{(j)}+f_jU_v^{(j)}\le\min\{f_j,q_j(F_v)\}$, and this certification threshold is at most $f_j$ and never
increases, since facilities are only added. 

We divide the counted certifications into three cases as in the uniform case.

\noindent\emph{(1) Multiplicative progress.}
Suppose that, immediately before a counted certification,
$\min\{f_j,q_j(F_v)\}>f_j/n$ and
$\xi_v^{(j)}\le\min\{f_j,q_j(F_v)\}/\rho$.
The certification then decreases the threshold by a factor
of at least $\rho$, because
\[
 \min\{f_j,q_j(F_v^+)\}\le\xi_v^{(j)}
 \le\frac{\min\{f_j,q_j(F_v)\}}{\rho}.
\]
Since the threshold starts at most $f_j$ and never increases,
before the $r$th certification of this kind we have
\[
 \frac{f_j}{n}<\min\{f_j,q_j(F_v)\}
 \le\frac{f_j}{\rho^{r-1}}
 \quad\Longrightarrow\quad
 r<1+\frac{\log n}{\log\rho}.
\]
Thus at most $1+\log n/\log\rho$ counted certifications
belong to this case.

\noindent\emph{(2) Small certification threshold.}
Suppose instead that $\min\{f_j,q_j(F_v)\}\le f_j/n$.
Conditional on $\mathcal H_v$, the threshold is fixed and
$U_v^{(j)}$ is still uniform. Therefore,
\[
 \Prb((v,j)\text{ is regularly certified}\mid\mathcal H_v)
 \le\frac{[\min\{f_j,q_j(F_v)\}-\xi_v^{(j)}]_+}{f_j}
 \le\frac1n.
\]
The first inequality above follows because regular certification requires
$f_jU_v^{(j)}\le\min\{f_j,q_j(F_v)\}-\xi_v^{(j)}$ and 
because $U_v^{(j)}$ is uniform. Moreover, certification also requires
$Z_v^{(j)}<\infty$, so its probability can only be smaller.
For the second inequality, this follows from
$\xi_v^{(j)}\ge0$ and $\min\{f_j,q_j(F_v)\}\le f_j/n$. Thus each client contributes at most $1/n$ in expectation to this case, and so summing over all clients of $C$,
even those whose certifications are not first in their batch,
bounds the expected count by
$\sum_{v\in C}1/n=|C|/n\le1$.

\noindent\emph{(3) Remaining certifications.}
For every remaining counted certification,
$\xi_v^{(j)}>\min\{f_j,q_j(F_v)\}/\rho$, and hence
\[
 f_jU_v^{(j)}
 \le\min\{f_j,q_j(F_v)\}-\xi_v^{(j)}
 <(\rho-1)\xi_v^{(j)}.
\]
We notice that every remaining counted certification belongs to a \emph{suspect} batch, i.e., a batch containing $v\in C_b$ with
$Z_v^{(j)}<\infty$ and
$f_jU_v^{(j)}<(\rho-1)\xi_v^{(j)}$. We let $N$ count these batches, so that via union bound
\begin{align*}
 \E[N]
 \le\sum_{v\in C}
   \Prb\left(f_jU_v^{(j)}<(\rho-1)\xi_v^{(j)}\right) =\sum_{v\in C}\min\left\{1,
       \frac{(\rho-1)\xi_v^{(j)}}{f_j}\right\}
 \le\frac{(\rho-1)\sum_{v\in C}\xi_v^{(j)}}{f_j}.
\end{align*}
Now, let us write $Q_1^{(j)},\ldots,Q_N^{(j)}$ for the keys of the
suspect batches, and let $\pi_1,\ldots,\pi_N$ be their arrival
order. Conditional on $N$ and these keys, which are distinct
almost surely, $\pi$ remains a uniformly random permutation.
By \Cref{cl:nonuniform-level-records}, every batch counted in
this case is a lower record among the suspect batches, and hence
\begin{align*}
 \E\!\left[\text{number counted in this case}
       \;\middle|\;Q_1^{(j)},\ldots,Q_N^{(j)}\right]
 &\le\sum_{r=1}^{N}
   \Prb\!\left(
      Q_{\pi_r}^{(j)}=\min_{\ell\le r}Q_{\pi_\ell}^{(j)}
      \;\middle|\;Q_1^{(j)},\ldots,Q_N^{(j)}\right)\\
 &=\sum_{r=1}^{N}\frac1r=H_N.
\end{align*}
Therefore, taking expectations and applying Jensen's inequality yields
\begin{align*}
 \E[\text{number counted in this case}]
 \le\E[H_N]&\le1+\log(1+\E[N]) \le1+\log\left(1+
       \frac{(\rho-1)\sum_{v\in C}\xi_v^{(j)}}{f_j}\right).
\end{align*}
Adding the bounds for the three cases gives
\[
 \E[L_C^{(j)}]
 \le3+\frac{\log n}{\log\rho}
 +\log\left(1+
       \frac{(\rho-1)\sum_{v\in C}\xi_v^{(j)}}{f_j}\right). \qedhere
\]
\end{proof}

The proof of \Cref{prop:short-nonuniform-records} immediately follows by applying \Cref{cl:nonuniform-level-records,cl:nonuniform-level-regular}
and use \Cref{cl:nonuniform-level-correspondence} to obtain $\sum_{v\in C}\xi_v^{(j)}
\le2\sum_{v\in C}d(c,v)=2\conn(C)$.

\subsection{Summing the prices}
\label{subsec:expanded-nonuniform-completion}

We now conclude the proof of \Cref{thm:nonuniform-batched}.

\begin{proof}[Proof of \Cref{thm:nonuniform-batched}]
Fix a comparison cluster $C$. First, we have that the harmonic bound \eqref{eq:nonuniform-level-harmonic} yields
\[
 \E[W_C]
 =\sum_{j\le j_C}f_j\E[K_C^{(j)}]
 \le H_B\sum_{j\le j_C}f_j
 \le2f_CH_B.
\]
For the other bound, we first account for the
nonregular certifications. If $\xi_v^{(j)}<f_j$, certification
gives $q_j(F_v^+)\le\xi_v^{(j)}<f_j$, after which no shifted
key above $f_j$ can be certified at this level. Thus there is
at most one such nonregular certification per level.
All other nonregular certifications satisfy
$f_j\le\xi_v^{(j)}\le2d(c,v)$. Since each charge in $W_C$
is covered by either a regular or a nonregular certification,
the geometric prices give
\begin{align*}
 W_C
 \le\sum_{j\le j_C}f_jL_C^{(j)}
   +\sum_{v\in C}\sum_{\substack{j\le j_C\\f_j\le2d(c,v)}}f_j
   +\sum_{j\le j_C}f_j \le\sum_{j\le j_C}f_jL_C^{(j)}+4\conn(C)+2f_C.
\end{align*}
Moreover, setting $\rho=\exp(\max\{1,\sqrt{\log(n+1)}\})$ and using $\log(1+x)\le x$,
\begin{align*}
 \log\left(1+\frac{2(\rho-1)\conn(C)}{f_j}\right)
 &\le\log\left(
       \rho\,\frac{f_C}{f_j}
       \left(1+\frac{2\conn(C)}{f_C}\right)\right)\\
 &\le\max\{1,\sqrt{\log(n+1)}\}+\log\frac{f_C}{f_j}
       +\frac{2\conn(C)}{f_C},
\end{align*}
since $f_j\le f_C$.
Finally, since opening costs are rounded to powers of two, they satisfy $\sum_{j\le j_C}f_j<2f_C$ and $\sum_{j\le j_C}f_j\log(f_C/f_j) \le f_C\log2\sum_{k\ge0}k2^{-k}=2f_C\log2$. Substituting all of the above three these inequalities into
\eqref{eq:short-level-bound} gives
\begin{align*}
 \E[W_C]
 &\le4\conn(C)+2f_C+\sum_{j\le j_C}f_j\E[L_C^{(j)}]\\
 &\le8\conn(C)
       +2f_C\left(4+\frac{\log n}{\max\{1,\sqrt{\log(n+1)}\}}+\max\{1,\sqrt{\log(n+1)}\}+\log2\right)\\
 &\le8\conn(C)+16\max\{1,\sqrt{\log(n+1)}\} f_C.
\end{align*}
Applying \Cref{prop:nonunif-reduction}, summing over the optimal
clusters, and using $\sum_Cf_C<2\OPT_{\mathrm{open}}$, we obtain
\begin{align*}
 \E[\ALG]
 &\le\sum_C\bigl(34\conn(C)+12\E[W_C]\bigr)\\
 &\le130\OPT_{\mathrm{conn}}
       +48\min\{H_B,8\max\{1,\sqrt{\log(n+1)}\}\}\OPT_{\mathrm{open}},
\end{align*}
which proves the first assertion.
For the second assertion, we use the two-copy algorithm of
\Cref{lem:joint-n-minus-B-transfer}. On the one hand, the singleton copy has the same law as one shuffled batch, so the bound with $B=1$ applies. On the other hand, The nonsingleton
copy has at most $2(n-B)$ clients and $B$ batches.
We, thus, apply the preceding cluster bound to both copies, using the
same optimal solution restricted to their clients. Their
connection costs add to $\OPT_{\mathrm{conn}}$, and each
opening cost is counted at most twice, so
\[
 \E[\ALG_{\mathrm{combined}}]
 \le O(1)\OPT_{\mathrm{conn}}
 +O\!\left(\min\left\{
       \log(B+1),\sqrt{\log(n-B+2)}
       \right\}\right)\OPT_{\mathrm{open}},
\]
proving the second assertion.
\end{proof}

\section{Lower bounds}
\label{lower:section}

We use Fotakis's binary-tree construction~\cite{Fotakis2008}, with more
batches at shallower depths so that random batch order still reveals many
successively deeper vertices. All opening costs are one, and facilities
may be opened at any vertex of the known tree. The bounds hold even if
each arriving batch is revealed in full and all batch sizes are known in
advance.

\begin{lemma}[Charging changes of the deepest revealed vertex]
\label{lower:record-charge}
Choose a uniform root-to-leaf path in a complete binary metric tree.
Each block consists of clients at one path vertex; block sizes, depths,
and order are independent of the path. Suppose each nonroot block's size
times its parent-edge length is at least one. If $R$ vertices are deeper
than every previously revealed vertex, counting the first reveal, then
$\E[\ALG]\ge\E[R]/3$. The block depths and any hierarchy independent
of the path may be known in advance.
\end{lemma}
\begin{proof}
Call these vertices \emph{records}, and write $T_v$ for the subtree rooted
at $v$. Fix the block order and the algorithm's randomness. Consider
consecutive records $v_a,v_b$, at depths $a<b$, just before $v_b$ is
revealed. If no facility is open in $T_{v_a}$, serving the earlier block
at $v_a$ cost at least one in connections across its parent edge. This
case is impossible when $v_a$ is the root, since a facility must already
be open. Otherwise, choose one open facility in $T_{v_a}$. Conditional on
the history, $v_b$ is uniform among its $2^{b-a}$ possible descendants:
earlier clients reveal no path vertex below $v_a$. Thus the chosen
facility lies outside $T_{v_b}$ with probability at least $1/2$; charge
its opening on this event 
.

No cost is charged twice: each record block is used once, and an excluded
facility lies outside every later record's subtree. Consequently
$\E[\ALG]\ge(\E[R]-1)/2$. Together with $\ALG\ge1$, this gives
$\E[\ALG]\ge\E[R]/3$.
\end{proof}

\begin{lemma}[A hard batched instance]
\label{lower:construction}
For every integer $h\ge4$, there is a distribution with
$B_h=2^{h+1}-1$ batches and $n_h-B_h=\exp(\Theta(h^2))$ such that
$\E[\ALG]\ge(h+1)/6$ and $\OPT\le5/4$.
\end{lemma}
\begin{proof}
In Fotakis's notation, take a complete binary tree $T$ of height $h$,
with edge length $D/m^i$ from depth $i$ to $i+1$. Set
$m=2^{h+3}$ and $D=1/m$, and choose a uniform path $v_0,\ldots,v_h$.
At $v_i$, put $2^{h-i}$ batches, each containing $m^i$ clients.
Opening only at $v_h$ gives
\[
 \OPT\le1+\sum_{i=0}^{h-1}2^{h-i}m^i
                   \sum_{\ell=i}^{h-1}\frac{D}{m^\ell}
 \le1+\frac{2^{h+1}-2}{m-1}<\frac54.
\]
For every nonroot batch, its size times its parent-edge length is
$m^iD/m^{i-1}=mD=1$. Moreover, $v_i$ is a record exactly when one of
its batches precedes every deeper batch. Hence
\[
 \Prb(v_i\text{ is a record})
 =\frac{2^{h-i}}{\sum_{\ell=i}^h2^{h-\ell}}\ge\frac12,
 \qquad
 \E[\ALG]\ge\frac{\E[R]}3\ge\frac{h+1}{6},
\]
where we apply \Cref{lower:record-charge}. Finally,
\[
 B_h=\sum_{i=0}^h2^{h-i}=2^{h+1}-1,
 \qquad
 m^h\le n_h=\sum_{i=0}^h2^{h-i}m^i<2m^h.
\]
Since $\log m=\Theta(h)$, these formulas give
$\log(n_h-B_h)=\Theta(h^2)$.
\end{proof}

\begin{theorem}[Batched lower bound for every $n$ and $B$]
\label{lower:joint}
For every $n\ge1$ and $1\le B\le n$, every randomized algorithm has
competitive ratio at least
\[
 \Omega\!\left(\min\!\left\{\log(B+1),
                 \sqrt{\log(n-B+2)}\right\}\right).
\]
This holds with uniform costs, and therefore also with nonuniform costs,
even if each arriving batch is revealed in full.
\end{theorem}
\begin{proof}
The construction must fit both the batch budget $B$ and the excess-client
budget $n-B$, and thus choose
\[
 h=\left\lfloor\frac12\min\!\left\{\log_2(B+1),
                  \sqrt{\log_2(n-B+2)}\right\}\right\rfloor.
\]
If $h<4$, the claimed factor is constant and follows from
$\ALG\ge\OPT$. Otherwise, the preceding construction satisfies
\[
 B_h\le2^{2h}-1\le B,
 \qquad
 n_h-B_h<2m^h\le2^{2h^2}\le\sqrt{n-B+2}\le n-B.
\]
Here $2m^h=2^{1+h(h+3)}\le2^{2h^2}$ for $h\ge4$.
Add $B-B_h$ singleton batches at the root, and add
$(n-B)-(n_h-B_h)$ clients to an existing root batch. This gives exactly
$n$ clients and $B$ batches. Opening also at the root keeps
$\OPT\le9/4$.

The added clients reveal nothing about the path. The original batches
retain their uniform relative order, so all their nonroot records remain
records, giving $\E[R]\ge h/2$. The hypothesis of
\Cref{lower:record-charge} is unchanged, and thus
$\E[\ALG]\ge h/6$. Averaging over the hidden path yields a fixed instance
with expected ratio $\Omega(h)$ for any randomized algorithm.
\end{proof}

\section*{Acknowledgments}
Fabrizio Grandoni is partially supported by the SNF Grants 200021-200731 and 200021-236706. Stefano Leonardi is partially supported by AI-PHOQUS PN RIC 2021-2027, co-financed by EU Italian Ministry MIMIT and MUR.
The authors thank Anupam Gupta and Michał Pawłowski for fruitful discussions in the early stages of this work. The authors also acknowledge using GPT-5.6 and GPT-6 Pro in developing this work, specifically to investigate proof strategies and support verification. The authors take full responsibility for the content of this work.

\bibliographystyle{alpha}
\bibliography{references}

\newcommand{\etalchar}[1]{$^{#1}$}
\begin{thebibliography}{AABC{\etalchar{+}}24}

\bibitem[AABC{\etalchar{+}}24]{AbbasiEtAl2024}
Fateme Abbasi, Marek Adamczyk, Miguel Bosch-Calvo, Jaros{\l}aw Byrka, Fabrizio Grandoni, Krzysztof Sornat, and Antoine Tinguely.
\newblock An {$O(\log\log n)$}-approximation for submodular facility location.
\newblock In {\em 51st International Colloquium on Automata, Languages, and Programming (ICALP 2024)}, volume 297 of {\em Leibniz International Proceedings in Informatics (LIPIcs)}, pages 5:1--5:20. Schloss Dagstuhl -- Leibniz-Zentrum f{\"u}r Informatik, 2024.

\bibitem[BGRS13]{ByrkaEtAl2013}
Jaros{\l}aw Byrka, Fabrizio Grandoni, Thomas Rothvo{\ss}, and Laura Sanit{\`a}.
\newblock {Steiner} tree approximation via iterative randomized rounding.
\newblock {\em Journal of the ACM}, 60(1):6:1--6:33, 2013.

\bibitem[CGLS23]{CohenAddadEtAl2023}
Vincent {Cohen-Addad}, Fabrizio Grandoni, Euiwoong Lee, and Chris Schwiegelshohn.
\newblock Breaching the 2 {LMP} approximation barrier for facility location with applications to {$k$}-median.
\newblock In {\em Proceedings of the 2023 Annual ACM-SIAM Symposium on Discrete Algorithms (SODA)}, pages 940--986. SIAM, 2023.

\bibitem[FGG{\etalchar{+}}25]{FotakisEtAl2025}
Dimitris Fotakis, Evangelia Gergatsouli, Themistoklis Gouleakis, Nikolas Patris, and Thanos Tolias.
\newblock Improved bounds for online facility location with predictions.
\newblock {\em Proceedings of the AAAI Conference on Artificial Intelligence}, 39(25):26973--26981, 2025.

\bibitem[Fot08]{Fotakis2008}
Dimitris Fotakis.
\newblock On the competitive ratio for online facility location.
\newblock {\em Algorithmica}, 50(1):1--57, 2008.

\bibitem[GKL22]{GuptaKehneLevin2021}
Anupam Gupta, Gregory Kehne, and Roie Levin.
\newblock Random order online set cover is as easy as offline.
\newblock In {\em Proceedings of the 62nd Annual IEEE Symposium on Foundations of Computer Science (FOCS 2021)}, pages 1253--1264. IEEE, 2022.

\bibitem[GKL24]{GuptaKehneLevin2024}
Anupam Gupta, Gregory Kehne, and Roie Levin.
\newblock Set covering with our eyes wide shut.
\newblock In {\em Proceedings of the 2024 Annual ACM-SIAM Symposium on Discrete Algorithms (SODA)}, pages 4530--4553. SIAM, 2024.

\bibitem[GMR26]{GuptaMolinaroRusso2026}
Anupam Gupta, Marco Molinaro, and Matteo Russo.
\newblock A learning perspective on random-order covering problems.
\newblock In {\em Proceedings of the 2026 SIAM Symposium on Simplicity in Algorithms (SOSA)}, pages 444--462. SIAM, 2026.

\bibitem[GS21]{GuptaSingla2021}
Anupam Gupta and Sahil Singla.
\newblock Random-order models.
\newblock In Tim Roughgarden, editor, {\em Beyond the Worst-Case Analysis of Algorithms}, pages 234--258. Cambridge University Press, 2021.

\bibitem[HJ26]{HuangJiang2026}
Yichen Huang and Shaofeng H.-C. Jiang.
\newblock The power of arrival times in random-order online facility location.
\newblock arXiv:2607.10564, 2026.

\bibitem[JLL{\etalchar{+}}22]{JiangEtAl2022}
Shaofeng H.-C. Jiang, Erzhi Liu, You Lyu, Zhihao~Gavin Tang, and Yubo Zhang.
\newblock Online facility location with predictions.
\newblock In {\em International Conference on Learning Representations (ICLR)}, 2022.

\bibitem[KNR23]{KaplanNaoriRaz2023}
Haim Kaplan, David Naori, and Danny Raz.
\newblock Almost tight bounds for online facility location in the random-order model.
\newblock In {\em Proceedings of the 2023 Annual ACM-SIAM Symposium on Discrete Algorithms (SODA)}, pages 1523--1544. SIAM, 2023.

\bibitem[Lan18]{Lang2018}
Harry Lang.
\newblock Online facility location against a {$t$}-bounded adversary.
\newblock In {\em Proceedings of the 29th Annual ACM-SIAM Symposium on Discrete Algorithms (SODA)}, pages 1002--1014. SIAM, 2018.

\bibitem[Li13]{Li2013}
Shi Li.
\newblock A 1.488 approximation algorithm for the uncapacitated facility location problem.
\newblock {\em Information and Computation}, 222:45--58, 2013.

\bibitem[LPS26]{LattanziPanigrahiSvensson2026}
Silvio Lattanzi, Debmalya Panigrahi, and Ola Svensson.
\newblock Online rounding and learning augmented algorithms for facility location.
\newblock In {\em International Conference on Learning Representations (ICLR)}, 2026.

\bibitem[LS04]{LeonardiSchaefer2004}
Stefano Leonardi and Guido Sch{\"a}fer.
\newblock Cross-monotonic cost sharing methods for connected facility location games.
\newblock {\em Theoretical Computer Science}, 326(1--3):431--442, 2004.

\bibitem[Mey01]{meyerson2001online}
Adam Meyerson.
\newblock Online facility location.
\newblock In {\em Proceedings of the 42nd Annual IEEE Symposium on Foundations of Computer Science (FOCS)}, pages 426--431. IEEE, 2001.

\bibitem[MYZ06]{MahdianYeZhang2006}
Mohammad Mahdian, Yinyu Ye, and Jiawei Zhang.
\newblock Approximation algorithms for metric facility location problems.
\newblock {\em SIAM Journal on Computing}, 36(2):411--432, 2006.

\bibitem[PT26]{PengTang2026}
Bo~Peng and Zhihao~Gavin Tang.
\newblock Random-order online facility location beyond uniform opening costs.
\newblock arXiv:2607.22496, 2026.

\bibitem[STA97]{ShmoysTardosAardal1997}
David~B. Shmoys, {\'E}va Tardos, and Karen Aardal.
\newblock Approximation algorithms for facility location problems.
\newblock In {\em Proceedings of the 29th Annual ACM Symposium on Theory of Computing (STOC)}, pages 265--274. ACM, 1997.

\end{thebibliography}

\end{document}